\documentstyle[prb,aps,floats,epsf,here]{revtex}

\begin{document}

\twocolumn[\hsize\textwidth\columnwidth\hsize
\csname@twocolumnfalse\endcsname
\draft
\title{Collective  modes  in  the  electronic  polarization  of
double-layer systems in the superconducting state}
\author{F. Forsthofer, S. Kind, J. Keller}
\address{Institute for Theoretical Physics, University of
Regensburg, D-93040 Regensburg, Germany}
\date{submitted to PRB, October 5, 1995}
\maketitle

\begin{abstract}
Standard weak coupling methods are used to study  collective modes
in the superconducting state of a double-layer system with intralayer
and interlayer interaction, as well as a Josephson-type coupling
and single particle hopping between
the layers by
calculating the electronic polarization function perpendicular to the
layers.
New analytical
results are derived  for  the  mode  frequencies corresponding
to fluctuations of the relative phase and amplitude of
the layer order
parameters in the case of interlayer
pairing and finite hopping $t$.  A new effect is found for finite
$k$-dependent  hopping:  then  the  amplitude  and  phase
fluctuations  are coupled.   Therefore  two collective  modes may
appear in the dynamical  c-axis conductivity  below the threshold
energy for breaking Cooper pairs. With help of numerical
calculations  we investigate  the temperature  dependence  of the
collective  modes and show how a plasmon corresponding to charge
fluctuations between the layers evolves  in the normal state.
\end{abstract}

 ] \narrowtext

%%%%%%%%%%%%%%%%%%%%%%%%%%%%%%%%%%%%%%%%%%%%%%%%%%%%%%%%%%%%%%%
%%%%%%%%%%%%%%%%%%%%%%%%%%%%%%%%%%%%%%%%%%%%%%%%%%%%%%%%%%%%%%%

\section{Introduction}

The problem of collective modes in superconductors  is a very old
one.   Already  Bogoliubov and Anderson \cite{AndBogol,Schrieffer}
pointed  out that charge oscillations  can couple to oscillations
of  the phase of the   superconducting   order   parameter
  via  the   pairing
interaction.  In a neutral system this would lead to a sound-like
collective  mode.  In a charged system the frequency of this mode
is  pushed  up to the  plasma  frequency  due  to the  long-range
Coulomb interaction \cite{Nambu,Parks},
  and at these high frequencies  this mode is
of no importance for the superconducting properties.

The situation is different  for modes which do not couple
to long-range  density  fluctuations. Leggett \cite{Leggett}
showed that in a two-band  superconductor  oscillations  in the
occupation   difference   between   the  two  bands   couple   to
phase fluctuations  of the  order  parameters  for the two  bands
giving  rise  to  a  collective   mode  with  a  frequency  below
$2\Delta$,  the threshold  energy for breaking Cooper  pairs.
Also oscillations of the amplitude of the order parameter are not
perturbed by charge fluctuations. The frequency of this mode,
however, is at the threshold for particle-hole  excitations
in normal isotropic superconductors.   Only in
special cases overdamping of this mode can be avoided \cite{Varma}.
Low-frequency  collective modes may also exist in superconductors
with a multi-component  order  parameter  \cite{Hirsch,Monien}.
Such order parameters are frequently discussed candidates for
some heavy fermion superconductors.

Another possibility  to avoid the influence of long-range Coulomb
forces   on  the  collective   modes  is  realized   in  strongly
anisotropic superconductors  \cite{Woelfle},  in particular in a
periodic system of superconducting  layers \cite{Fertig}.  In the
case of a finite value of the wave-vector $q_\perp$ perpendicular
to the layers density fluctuations within the layers do not
build up long-range Coulomb forces. In the limit $\vec q \to 0$ only
the Coulomb interaction between the layers remains.

Recently  the question  of collective  modes has been brought  up
again  in connection  with the multi-layer  structure  of the
high-T$_c$ superconductors.
   In a double-layer  system  like BSCCO or
YBCO one obtains two electronic bands for the motion of electrons
parallel to the layers corresponding  to states with symmetric or
antisymmetric  wave functions \cite{OkAndersen}.
  In the superconducting  state both
bands  acquire  gaps.  In such systems  one can discuss different
collective  modes \cite{Palistrant,WuGriffin,Kuboki} corresponding
to fluctuations  in the occupation number of the two bands and to
charge  oscillation   between  the  two  layers.

In  a series  of  papers
\cite{Gerhard1,Gerhard2} we  have
investigated  in  detail  the
electronic polarization of double-layer systems in the normal and
superconducting  state and have studied  the influence  of charge
fluctuations  between  the  layers  on  the  renormalization   of
transverse  c-axis phonons.   Here we have assumed  a k-dependent
tight-binding   coupling   between   the   layers   and   pairing
interactions  for  electrons  within  the same  and in different
layers. In our numerical calculations \cite{Gerhard2} we included
both the vertex
corrections due to the BCS-interaction and the Coulomb
interaction between the layers,  thus  taking  into account  the
effect  of possible
collective  modes.  We found a shift of
phonon  frequencies  in the  superconducting  state  which  is in
reasonable  agreement  with FIR-experiments.   As in this case we
assumed  a fairly large value for the tight-binding  coupling  the
collective   modes   are   strongly   damped   by  quasi-particle
excitations.   In the present paper we analyze in more detail the
conditions  for  undamped  collective  modes  in  the  electronic
polarization.   The results depend crucially on the values of the
different pairing interactions.

In principle  one has to
distinguish  three  different  types  of  interactions:   1.   an
interaction between electronic densities within one layer, 2. an
interaction between electronic densities in different layers.  3.
an interaction  between mixed densities from the two layers.  All
three interactions  can be mediated by phonons (or other types of
bosons):  the first  two by a change  of local  potentials
induced  by
lattice  displacements,  the third one by a change of the hopping
energy  between  the  two  layers.  The  first  two  interactions
conserve  the number  of particles  in the two layers  while  the
third  interaction  may  interchange  particles  between  the two
layers.  In particular, this interaction allows a transfer of two
particles  like in a Josephson coupling.   This type of coupling,
which can also be derived from a second order hopping process has
gained much interest recently.  P.~W.~Anderson and coworkers have
argued  that in strongly  correlated  electron  systems  coherent
single-particle  hopping  between  the two layers  is suppressed,
while a coherent momentum conserving second-order hopping process
is  possible   \cite{AndersonSc}.    In  such  an
interlayer-tunneling  model without single-particle  hopping  two
collective  modes involving fluctuations  of the phase and
amplitude difference of the layer order-parameters are found
\cite{WuGriffin}.

In a two-layer  system  one  has to consider  two  order  parameters
corresponding  to  pairing  of  electrons  in one  layer  and  in
different  layers.   This  leads  in general  to different  order
parameters  (and  gaps)  for the two bands  which  may even  have
different  signs for s-wave pairing.   Depending  on the relative
sign of the band order-parameters one can distinguish  between two
different pairing types, which both have s-wave symmetry:  In the
case of dominant intralayer  interaction  the order parameter  of
the two bands  have equal  sign,  while for dominant  interlayer
interaction         they        have        opposite         sign
\cite{Combescot,Golubov,Mazin,Levin}. In Ref.
\cite{swave,Normand} it
is shown  that  antiferromagnetic  interactions  between  the two
layers favour a superconducting state with interlayer pairing and
anisotropic s-wave symmetry.

In the following  discussion  we will consider all three types of
pairing  interactions  mentioned  above  and will also take  into
account single-particle hopping between the layers. In this paper
we consider only s-wave pairing by neglecting the k-dependence of
the pairing interaction. We discuss the collective modes for both
pairing types and calculate the electronic  polarization  between
the  two  layers,  because  this  quantity  enters  directly  the
dynamical conductivity for electric field vectors in c-direction.
Our results extend earlier work on collective  modes in two-layer
systems mainly in the following respects:
1. we calculate collective modes in
the case of interlayer pairing, 2. we find that a
k-dependent tight-binding hopping between the layers
couples  amplitude  and phase modes, making it possible  that both
modes appear below the threshold  frequency in
the optical conductivity.

Our paper is organized  as follows:  In the following  section we
specify  our model and write down the interactions  in a $4\times
4$-Nambu matrix notation. Then we calculate the self-energies and
discuss the self-consistency  equations.  In section III we solve
the vertex equations for the polarization function in the neutral
and  charged  system.  The  collective  modes  for  k-independent
hopping  matrix element are studied  in section IV for intralayer
pairing and interlayer  pairing. In
section  V numerical  results for the polarization  function  are
presented. The results are summarized in section VI.

%%%%%%%%%%%%%%%%%%%%%%%%%%%%%%%%%%%%%%%%%%%%%%%%%%%%%%%%%%%%
\section{Nambu formalism for two-band systems}
\subsection{Model}

We consider an electronic  double-layer  system described  by the
Hamiltonian

\begin{eqnarray}
H_0 &=& \sum_{k\sigma} \epsilon_k
 (c^\dagger_{1k\sigma}c^{\phantom{\dagger}}_{1k\sigma}
+ c^\dagger_{2k\sigma}c^{\phantom{\dagger}}_{2k\sigma})
\nonumber
\\
&&{\phantom{xxxxxxxx}}
  + t_k  (c^\dagger_{2k\sigma}
c^{\phantom{\dagger}}_{1k\sigma}
 + c^\dagger_{1k\sigma} c^{\phantom{\dagger}}_{2k\sigma})
\end{eqnarray}
and the interaction
\begin{eqnarray}
H_S &=& {1\over 2} \sum_{k k'q \sigma \sigma'} \sum_i {\phantom+}
 V_\parallel \,
 c^\dagger_{i   k+q  \sigma}
 c^\dagger_{i  k'-q\sigma'}  c^{\phantom\dagger}_{i  k'\sigma'}
 c^{\phantom\dagger}_{i  k  \sigma}\nonumber
\\
&&{\phantom{ {1\over 2} \sum_{k k'q \sigma \sigma'}  } }
+V_\perp \,
 c^\dagger_{i   k+q  \sigma}
 c^\dagger_{j  k'-q\sigma'}  c^{\phantom\dagger}_{j  k'\sigma'}
 c^{\phantom\dagger}_{i  k  \sigma} \nonumber
\\
&&{\phantom{ {1\over 2} \sum_{k k'q \sigma \sigma'}  } }
+J\,
(
 c^\dagger_{i   k+q  \sigma}
 c^\dagger_{i  k'-q\sigma'}  c^{\phantom\dagger}_{j  k'\sigma'}
 c^{\phantom\dagger}_{j  k  \sigma} \nonumber
\\
&&{\phantom{ {1\over 2} \sum_{k k'q \sigma \sigma'}
 + J \,} }
 +c^\dagger_{i   k+q  \sigma}
 c^\dagger_{j  k'-q\sigma'}  c^{\phantom\dagger}_{i  k'\sigma'}
 c^{\phantom\dagger}_{j  k  \sigma}
)
\end{eqnarray}
Here $t_k$ describes  a tight-binding  coupling  between  the two
layers $i=(1,2), j=3-i$.
The couplings $V_\parallel$,  $V_\perp$,  $J$ are effective pairing
interactions.
$V_\parallel$   describes  the  interaction   between  electronic
densities  within one layer, $V_\perp$ between different  layers,
and $J$ is the coupling  between mixed densities.   The origin of
these interactions  is a combination of an attractive interaction
due to the exchange  of phonons  (or other bosons) and a screened
repulsive  Coulomb interaction.   Because $J$ includes  intrinsic
tunneling  of Cooper  paris from one layer  to the other,  $J$ is
called  Josephson   coupling.    Here  we  neglect  any  momentum
dependence of these interactions  (except of a cut-off introduced
later).

The  Hamiltonian $H_0$   can  be  diagonalized
 by  introducing   new
fermionic operators $a_{\alpha k\sigma}$
$$
a_{1k\sigma}   =  {1\over   \sqrt{2}}   (c_{2k\sigma}   -
c_{1k\sigma})
,\quad
a_{2k\sigma}   =  {1\over   \sqrt{2}}   (c_{2k\sigma}   +
c_{1k\sigma})
$$
corresponding  to states  with antisymmetric  and symmetric  wave
function  on the two layers (this  symmetry  is not  broken  by
introducing  the interactions).   We then  obtain  two bands  with
quasi-particle energies
$$
\epsilon_{1k}  = \epsilon_k  - t_k, \quad
\epsilon_{2k}   =  \epsilon_k  +  t_k
$$

In  order  to treat  superconductivity  it is useful  to combine
the Fermi operators to a Nambu spinor \cite{Schrieffer,Nambu}
$$
\Psi_k =\left(
                 a_{1k\uparrow}^{\phantom\dagger},\,\,\,
                  a^\dagger_{1-k\downarrow},\,\,\,
		  a_{2k\uparrow}^{\phantom\dagger},\,\,\,
                  a^\dagger_{2-k\downarrow}
         \right)^t
$$
With help of these spinors we can express the Hamiltonian  of the
two-layer system as follows (apart from constants):
\begin{equation}
H_0  =  \sum_k  \epsilon_k  \Psi^\dagger_k  D^{03}
\Psi^{\phantom\dagger}_k  - t_k
\Psi^\dagger_k D^{33}\Psi^{\phantom\dagger}_k
\end{equation}
\begin{eqnarray}
H_S  &=& {1\over  2}  \sum_{k,k',q}   V
(\Psi^\dagger_{k+q}  D^{03}
\Psi^{\phantom\dagger}_k)(\Psi^\dagger_{k'-q}  D^{03}
\Psi^{\phantom\dagger}_{k'})
\nonumber
\\
&&{\phantom{ {1\over  2}  \sum  } }
  + \bar V (\Psi^\dagger_{k+q}  D^{13} \Psi^{\phantom\dagger}_k)
(\Psi^\dagger_{k'-q}  D^{13} \Psi^{\phantom\dagger}_{k'})
\label{Hinteraction}
\\
&&{\phantom{ {1\over  2}  \sum  } }
  + J  (\Psi^\dagger_{k+q}  D^{33} \Psi^{\phantom\dagger}_k)
       (\Psi^\dagger_{k'-q}  D^{33} \Psi^{\phantom\dagger}_{k'})
\nonumber
\end{eqnarray}

Here $V= (V_\parallel + V_\perp)/2$,
$\bar V = (V_\parallel  - V_\perp)/2$.
$D^{jl}$  are $4\times4$  matrices  in Nambu space which are
constructed from two sets of Pauli matrices
$D^{jl} = \tau_j\sigma_l$.
Example:
$$
D^{13} = \tau_1 \sigma_3 = \left(\begin{array}{cc}
                                    0 & \sigma_3 \\
                                    \sigma_3 & 0
                                  \end{array}
                           \right)
$$
The terms containing the matrices $D^{03}, D^{33}$ describe the sum
and difference of density operators of the two bands, while the term
with $D^{13}$ contains fermionic operators from different bands.
Accordingly $\bar V$ leads to interband transitions while $V$ and $J$
cause intraband transitions only.
In principle one can derive a further interaction mediated by
phonons which couples
operators characterized by the matrices $D^{33}$ and $D^{03}$.
Such an interaction is
omitted here, because  it has no influence on the
interlayer polarization function.

 For the interactions $V,\bar V, J$  a cut-off  has to be
introduced  either in momentum space or frequency space.  We
introduce the cut-off in momentum space:
\begin{equation}
V,\bar V, J \left\{
       \begin{array}{ccc}
             \ne & 0 & {\rm if}\quad  |\epsilon_k-\mu|,
                           |\epsilon_{k'}-\mu| < \omega_c\\
              =  & 0 & {\rm otherwise}
      \end{array}
           \right .
\label{cutoff}
\end{equation}
$\omega_c$ is the cut-off energy, $\mu$ the
chemical potential.  Note that the cut-off has
been introduced for the
energy    $\epsilon_k$    and    not    for   the   band-energies
$\epsilon_{\alpha k}$.
Otherwise  we  obtain  inconsistencies   in  solving  the  vertex
equations.

In the following  we are primarily interested  in the calculation
of the interlayer polarization function which is the correlation
function  of the  electronic  polarization perpendicular  to the
layers.  The latter is described (up to a factor $ed/2$, where $d$ is
the layer separation and $e$ the electronic charge) by the operator
\begin{equation}
P = \sum_{k\sigma}
c^\dagger_{2k\sigma} c^{\phantom\dagger}_{2k\sigma}
 - c^\dagger_{1k\sigma} c^{\phantom\dagger}_{1k\sigma}
\end{equation}
There   are  three   other   operators   which   couple   to  the
interlayer polarization in the vertex equations. These are:
\begin{eqnarray}
\nonumber
\Phi &=& -i\sum_{k} c^\dagger_{2k\uparrow}
c^\dagger_{2-k\downarrow} - c^\dagger_{1k\uparrow}
c^\dagger_{1-k\downarrow} - c^{\phantom\dagger}_{2-k\downarrow}
c^{\phantom\dagger}_{2k\uparrow} +
c^{\phantom\dagger}_{1-k\downarrow}
c^{\phantom\dagger}_{1k\uparrow}
\\
\nonumber
A &=& \sum_{k} c^\dagger_{2k\uparrow}c^\dagger_{2-k\downarrow} -
c^\dagger_{1k\uparrow}c^\dagger_{1-k\downarrow} +
c^{\phantom\dagger}_{2-k\downarrow}c^{\phantom\dagger}_{2k\uparrow}
- c^{\phantom\dagger}_{1-k\downarrow}
c^{\phantom\dagger}_{1k\uparrow}
\\
j &=& - i \sum_{k\sigma} c^\dagger_{2k\sigma}
c^{\phantom\dagger}_{1k\sigma} -
c^\dagger_{1k\sigma}c^{\phantom\dagger}_{2k\sigma}
\end{eqnarray}
In  Nambu  space  a  $4\times  4$  matrix
corresponds  to each operator:
\begin{equation}
P^{jl} = \sum_k \Psi^\dagger _k D^{jl} \Psi^{\phantom\dagger}_k
\label{operators}
\end{equation}
where $P=P^{13}, \Phi = P^{12}, A=P^{11}, j = P^{20}$.

$\Phi$ and $A$ are  called  phase  and amplitude  operator  in the
literature \cite{WuGriffin},  because  in  the  case
that  the  phases  of the  layer
order-parameters  are zero $\Phi$ and $A$  measure the difference  of
the  phases   and  amplitudes   of  the  layer  order-parameters.
$j$  is  related  to  the  interlayer  current  density.   This  is
strictly  so  only  in the  case  of constant  interlayer hopping
$t_k=t$ and vanishing Josephson-coupling $J$; then the
equation of motion  $2t j = \dot  P$ is fulfilled.  In the  case
$J\ne0$ the current-density operator becomes more complicated, it
contains also two-particle operators.  Nevertheless $j$
will be called
current operator in the following.

\subsection{Self-consistency equations}

The Green's functions can be combined to a
$4\times 4 $-Nambu matrix.
We assume the Green's function matrix to be diagonal
in the band  indices:
\begin{equation}
G(k,z) =\left(
 \begin{array}{cc}
      G_1(k,z) & 0 \\
      0 & G_2(k,z)  \\
\end{array}\right)
\end{equation}
where $G_\alpha(k,z)$  are $2\times2$ matrices for each band.
This  assumption which neglects pairing in different bands is
reasonable  since  the  two  bands  (in  the
presence  of  a finite  tight-binding  coupling)  have  different
symmetries \cite{Levin},  and this  symmetry
is preserved  by the interaction
$H_S$. One consequence of this assumption is, that the phases
of the order parameters $<c_{ik\uparrow}c_{i-k\downarrow>}$
for pairing
in one layer are equal for the two layers
(equal amplitude  and phase).
Intraband pairing seems to be the ground
state  for  all  negative  values  of $J$,  while  it may  become
unstable  for large  positive  $J \gg \vert  t_k\vert$.   This is
supported  by  the  calculation  of  the  groundstate  of a model
with  two  electrons  on two  sites  with  all  the  interactions
contained   in   the   Hamiltonian   $H_0   +   H_S$.    Defining
operators $a_{\alpha\sigma}  = {1\over \sqrt{2}}
(c_{2\sigma}   \pm   c_{1k\sigma})$   for   the   single-particle
eigenstates  of $H_0$  one finds  that  the groundstate  contains
only diagonal terms
$a^\dagger_{\alpha\uparrow}a^{\dagger}_{\alpha\downarrow}\vert
0>$ as long
as the Josephson  coupling  is attractive  $J\le  0$ but contains
non-diagonal terms for large positive values  $J \gg \vert t_k\vert$.
In our numerical  calculations  of the polarization  function  we
will show that an instability occurs for large positive Josephson
coupling $J$.

The bare  Green's function is given by:
\begin{equation}
G^{-1}_{0 \alpha}(k,z) = z \sigma_0 - (\epsilon_{\alpha,k} - \mu)
\sigma_3
\end{equation}
In the presence of the pairing interaction  we have
self-energy corrections:
\begin{equation}
G^{-1}(k,z) = G_0^{-1}(k,z) - \Sigma(z)
\end{equation}
where  the  self-energy  $\Sigma(z)$  contains  contributions
from
intraband  and interband  interactions.
\begin{eqnarray}
\Sigma   &= & -  {1\over\beta}   \sum_{k   \omega_m}{\phantom{+}}
V  D^{03}
G(k,i\omega_m) D^{03} +  \bar V  D^{13}
G(k,i\omega_m) D^{13}
\nonumber
\\
&&
{\phantom{-  {1\over\beta}   \sum_{k   \omega_m}}}
 + J  D^{33} G(k,i\omega_m) D^{33}
\end{eqnarray}
The self-energy is diagonal
in the band indices
\begin{equation}
\Sigma =\left(  \begin{array}{cc}
         \Sigma_1 & 0 \\
         0 & \Sigma_2 \\
          \end{array}
        \right)
\end{equation}
and the $2\times2$ components $\Sigma_{\alpha}$ are given by
\begin{eqnarray}
\Sigma_1 &=& - {1\over\beta} \sum_{k \omega_m}  W
\sigma_3  G_1  (k,i\omega_m)  \sigma_3  +  \bar  V \sigma_3  G_2(
k,i\omega_m) \sigma_3
\nonumber
\\
\Sigma_2 &=& - {1\over\beta}  \sum_{k
\omega_m}    W  \sigma_3  G_2  (k,i\omega_m)  \sigma_3  +\bar  V
\sigma_3 G_1 (k,i\omega_m)  \sigma_3
\label{sigma12}
\end{eqnarray}
with the intraband interaction $ W=V+J$.
While the interactions $V$, $J$ enter in
the same  way into  the self-energy,  this is not so in the vertex
equations.  Therefore $J$ cannot simply be incorporated  into the
interaction V by a proper redefinition.

Writing $\Sigma_\alpha = -\Delta_\alpha \sigma_1 - x_\alpha \sigma_3$
we have
\begin{equation}
G^{-1}_{\alpha}(k,z) = z \sigma_0 - \xi_{\alpha k} + \Delta_\alpha
\sigma_1
\end{equation}
with $ \xi_{\alpha k} = \epsilon_{\alpha k}    +   x_\alpha   -\mu$.

$\Delta_\alpha$ is the energy gap of band $\alpha$.
 $x_\alpha$  is an energy  shift  due to the pairing
interaction  which  is also present  in the normal  state.
In a one-layer system $x_\alpha$ can be
incorporated into the chemical potential $\mu$,
but this is not possible in the two-layer
system with interlayer hopping.

Performing   the  frequency   summation   we  get  the  following
self-consistency equations for the gaps and energy shifts:

\begin{eqnarray}
\Delta_1 &=& - W \Delta_1 I_1 - \bar V \Delta_2 I_2
\nonumber
\\
\Delta_2 &=& - W \Delta_2 I_2 - \bar V \Delta_1 I_1
\label{gapeqn}
\end{eqnarray}
\begin{eqnarray}
x_1 &=& -  W K_1 - \bar V K_2
\nonumber
\\
x_2 &=& - W K_2 - \bar V K_1
\label{xeqn}
\end{eqnarray}
with the integrals
$$
I_\alpha = \sum_k {1\over {2 E_{\alpha k}}}
\tanh {\beta E_{\alpha k}\over 2},\quad
K_i = \sum_k {\xi_{\alpha  k}\over {2 E_{\alpha k}}} \tanh {\beta
E_{\alpha k}\over 2}
$$
and $ E_{\alpha k}= \sqrt{ \xi_{\alpha k}^2 + \Delta_\alpha^2}$

Without a cut-off the integrals $I_i$ would diverge.
Approximately  the integrals have the values:
\begin{equation}
I_\alpha \simeq N_\alpha \ln (2 \omega_c / Max(\Delta_\alpha, T))
\label{divergent}
\end{equation}

where $N_\alpha$ is the density  of states  of band $\alpha$ per
spin at the Fermi surface,
 and $\omega_c$
is the cutoff.

The  self-consistency equations (\ref{xeqn})
describing  the energy  shifts
$x_{1,2}$  of  the  two  bands  can  be solved  approximately  by
observing  that the integrands  are nearly  step functions.
Assuming constant
density of states $N_1$, $N_2$ (this is given by a
quadratic 2D-dispersion) and
$k$-independent interlayer hopping $t$ one finds
\begin{eqnarray}
x_1 &=& W N_1 (t+x_1) - \bar V N_2 (t-x_2)
\nonumber
\\
x_2 &=& - W N_2 (t-x_2) + \bar V N_1 (t+x_1)
\label{x0}
\end{eqnarray}
In the limit of $W,\bar V \to 0$ and cut-off
$\omega_c \to \infty$
these shifts vanish, but the ratio $(x_2-x_1)/(W-\bar  V)$
stays finite
($N_0 = (N_1+N_2)/2)$:
\begin{equation}
{x_2-x_1\over W-\bar V} \to - 2 N_0 t
\label{weakcoupling}
\end{equation}

The solution of the gap equations for a two-band system is discussed
already by Leggett \cite{Leggett}; the qualitative results can be
summarized as follows:
There exists no solution, if $W>0$ and $W^2>{\bar V}^2>0$;
then the intralayer
$W_\parallel=V_\parallel+J$ and the interlayer coupling
$W_\perp=V_\perp+J$ are  repulsive. There
are two nontrivial solutions, if $W<0$ and
$W^2>{\bar V}^2>0$;
 in this case $ W_\parallel$ and $W_\perp$
 are attractive.
The solution corresponding to the ground state depends on the sign
of the interband interaction
$\bar V$.  If $\bar V <0 $, the state where the gaps
have the same sign
($\Delta_1 \Delta_2> 0$) is
stable; if $\bar V>0$, then  the gaps have opposite signs
($\Delta_1 \Delta_2< 0$). The reason is a
term  in the free energy
 proportional to $\bar V \cos(\phi_1-\phi_2)$,
where $\phi_1, \phi_2$ are
the phases of the order parameters $\Delta_1, \Delta_2$.

If the  density  of states  for the two  bands  are equal
(this happens for a constant density of states
and for $k$-independent interlayer hopping $t$)
the two integrals  $I_i$ are equal for equal values of
the gaps ($I_1=I_2$ for $\Delta_1^2  = \Delta_2^2$)  then for all
solutions   of  the  gap  equations   one  always   finds  $\vert
\Delta_1\vert  = \vert \Delta_2\vert$ and $x_1=-x_2$
irrespective of the values
of the coupling  constants,  i.e.  one either has pure intralayer
pairing  for $W_\parallel<0$   with  $  (\Delta_1  +  \Delta_2)/2   =
\Delta_\parallel \neq 0, (\Delta_2 - \Delta_1)/2 = \Delta_\perp =
0$ or pure interlayer pairing for $W_\perp<0$ with $\Delta_\perp \neq
0$ and $\Delta_\parallel =0$.   In the
more  general  case the density  of states  of the two bands  are
different  at the  Fermi  surface,  and  the two  integrals  have
different values also in the case of equal or vanishing gaps, and
the two order parameters are coupled.

For the numerical  solutions  of the
self-consistency equations and vertex equations
with k-dependent hopping  we use
the following  simple model by choosing  two bands with quadratic
dispersion  but different effective masses (and different density
of states):
\begin{eqnarray}
\epsilon_{1k}  &=& \epsilon_k  - t_k = k^2/(2m_1) - t_0
\nonumber\\
\epsilon_{2k}  &=& \epsilon_k  + t_k = k^2/(2m_2) + t_0
\end{eqnarray}

In the numerical calculations  we use the following dispersion
 parameters  to get k-dependent hopping:
\begin{eqnarray}
&2m_1/\hbar^2=1\,\rm eV^{-1}\AA^{-2},\,
 2m_2/\hbar^2=1.2\,\rm eV^{-1}\AA^{-2},&
\nonumber
\\
& t_0=0,\,
\mu=0.3\,{\rm eV},\,\omega_c=0.25\,{\rm eV}&
\label{parameter}
\end{eqnarray}
For the intraband and interband coupling constants $W=V+J$ and
$\bar V$, entering the
self-consistency equations we choose
\begin{equation}
N_0 W = -0.139,\,\,\, N_0 \bar V = \pm 0.185
\label{WbV}
\end{equation}
\begin{figure}[h]
\begin{center}
\leavevmode
\epsfxsize=0.95\linewidth
\epsfbox{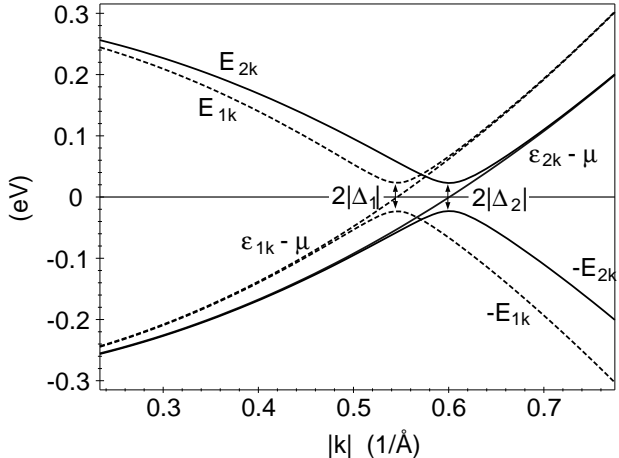}
\caption{\label{dispm1=1m2=1.2}Dispersions in the normal state
$\epsilon_{1k}$, $\epsilon_{2k}$ and quasi-particle dispersions in
the superconducting state (T=0) in the $k$-region
$|\epsilon_k|< \omega_c$. The  parameters (22,23) are used.
}
\end{center}
\end{figure}
\begin{figure}[h]
\begin{center}
\leavevmode
\epsfxsize=0.85\linewidth
\epsfbox{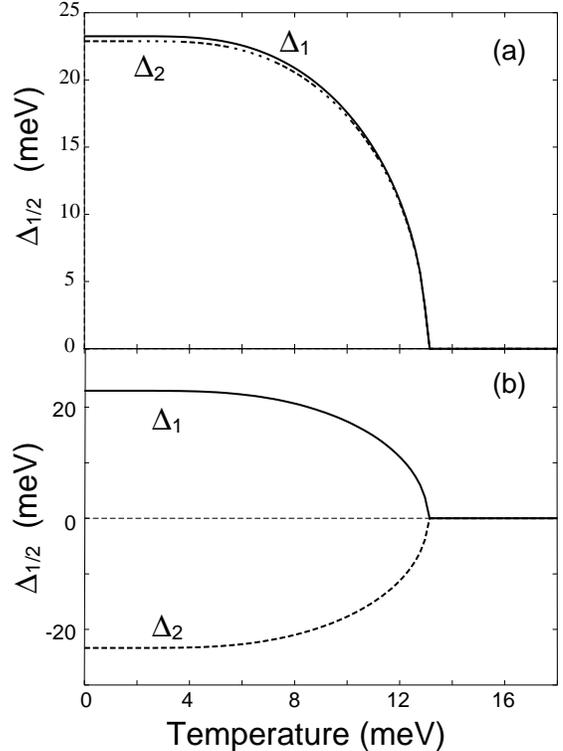}
\caption{\label{gapstempII1} Gaps  as
function of temperature for negative  (a)
and positive interband coupling $\bar V$ (b).
The parameters (22,23)
are used.
}
\end{center}
\end{figure}
We  have  chosen   these  values   for  the  coupling   constants
in order to get $T_c$-values
which  are
appropriate for high-T$_c$ superconductors and ensure also the
appearance of the collective modes we want to study.

Fig. \ref{dispm1=1m2=1.2} shows the dispersion for
the two bands in
the normal state and the quasi-particle dispersions in the
superconducting state at zero temperature
in the k-region $|\epsilon_k-\mu| <\omega_c$.
Figures \ref{gapstempII1}(a), \ref{gapstempII1}(b)
show the temperature dependence of the gaps
for negative $\bar V$ (a) and positive $\bar V$ (b).
Because the effective masses $m_1$ and $m_2$ are
nearly the same (\ref{parameter}) the
values of the two gaps $\Delta_1$ and $\Delta_2$ are nearly
the same.  For repulsive
interband coupling $\bar V$ the stable  gaps have different signs.
In both cases intralayer and interlayer pairing
are mixed. But for negative $\bar V$ intralayer pairing
dominates, for positive $\bar V$ interlayer pairing dominates.
$T_c$ is 155 K.
We have also calculated the level shifts $x_i$. They are
practically temperature independent.

%%%%%%%%%%%%%%%%%%%%%%%%%%%%%%%%%%%%%%%%%%%%%%%%%%%%%%%%%%%
\section{ Vertex equations}

Now we will set up the integral equations for the vertex functions
corresponding  to the interlayer polarization and related
operators defined in (\ref {operators}) for the neutral and
charged system.  We are interested  in optical response functions
for electric field vectors perpendicular  to the layers.  Here we
can assume that the  external wave vector
is zero.

\subsection{Neutral system}

 We consider response functions where the
bare vertex is one of the matrices $D^{jl}$,
corresponding to the operators
(\ref {operators}).
 In the standard ladder approximation  the vertex equation  for the
renormalization of a vertex with matrix $D^{jl}$ reads
\begin{eqnarray}
\lefteqn{
\Gamma^{jl}(i\omega_s)     =D^{jl} }
\nonumber\\
 &&     -    {1\over   \beta}
\sum_{k\omega_n}{\phantom{+}}
V   D^{03}    G(k,i\omega_n+i\omega_s)    \Gamma^{jl}(i\omega_s)
G(k,i\omega_n)  D^{03}
\nonumber
\\ &&
{\phantom{  -    {1\over   \beta} \sum_{k\omega_n}} }
+ \bar V D^{13}  G(k,i\omega_n+i\omega_s)
\Gamma^{jl}(i\omega_s)G(k,i\omega_n) D^{13}
\nonumber
\\ &&
{\phantom{   -    {1\over   \beta} \sum_{k\omega_n}} }
 + J D^{33}  G(k,i\omega_n+i\omega_s)
\Gamma^{jl}(i\omega_s)G(k,i\omega_n) D^{33}
\label{vertexeqn1}
\end{eqnarray}
The vertex function depends only on the external frequency
$\omega_s$, not on the momentum $k$ (apart from
the cut-off) and the internal frequency of the Green's function,
because the
interactions are assumed to be constant.
The label  $jl$
indicates  to which bare vertex the vertex function  belongs. In
the  case  of the polarization  we have  to
consider the matrix $D^{13}$.  Of
course,   due   to  the  interactions   the   renormalized   vertex
$\Gamma^{jl}$ contains also contributions from other matrices.

The interactions  $V,J$  induce  only  intraband transitions,  the
interaction $\bar V$ only interband transitions. These interactions
do  not  change  the  off-diagonal  character
of  the  bare vertices,
therefore  we can make the following  ansatz  for the renormalized
vertex of the neutral system
\begin{equation}
\Gamma^{jl} = \left(
             \begin{array}{cc}
                        0 & \gamma \\
                        \hat \gamma & 0\\
            \end{array}
             \right)
\end{equation}
with $2\times 2$ matrices
$\gamma(i\omega_s), \hat\gamma(i\omega_s)$.
Then  the  system  of  vertex  equations  can  be written  as two
coupled equations for $2\times 2$ matrices:
\begin{eqnarray}
\gamma + W' B + \bar V \hat B &=& I
\nonumber \\
\hat \gamma + W' \hat B + \bar V B &=& \hat I
\end{eqnarray}
where  $W'  =  V-J=(V_\parallel+V_\perp)/2-J$.  The  $2\times  2$
matrices  $I$ and $\hat  I$ are the upper  right  and lower  left
part of the matrix $D^{jl}$.
In the case of the polarization $D^{13}$
we have $I=\sigma_3$, $\hat I=-\sigma_3$.
The quantites $B$ are defined as
$$
B(i\omega_s)  = {1\over  \beta}  \sum_{\omega_n}\sum_k  \sigma_3
G_1(k,i\omega_n+i\omega_s)   \gamma(i\omega_s)  G_2(k,i\omega_n)
\sigma_3
$$
\begin{equation}
\hat B(i\omega_s)  ={1\over  \beta}
\sum_{\omega_n}\sum_k  \sigma_3
G_2(k,i\omega_n+i\omega_s) \hat \gamma (i\omega_s)
G_1(k,i\omega_n)
\sigma_3
\label{B}
\end{equation}
depending linearly on $\gamma(i\omega_s)$.
Note, that in contrast to the self-consistency equations
(\ref{gapeqn}, \ref{xeqn}) ($W=V+J$) the
interaction $J$ enters here with an other sign ($W'=V-J$).

For the solution  of this system of equations it is convenient  to
decompose $\gamma, \hat \gamma$  and also $B, \hat B$
into Pauli matrices:
\begin{eqnarray}
\gamma  = \gamma_0  \sigma_0  + \gamma_1  \sigma_1  + \gamma_2  i
\sigma_2 + \gamma_3 \sigma_3
\nonumber
\\
B = B_0 \sigma_0 + B_1 \sigma_1 + B_2 i\sigma_2 + B_3 \sigma_3
\label{Paulientw}
\end{eqnarray}
If we write the coefficient functions $\gamma_i$ as vector
$\underline {\gamma} = (\gamma_0, \gamma_1, \gamma_2, \gamma_3)$
we can express the linear dependence of $B$ on
these coefficients in matrix form:
\begin{equation}
\underline {B} = K \underline {\gamma}, \quad \underline {\hat B}
= \hat K \underline {\hat \gamma}
\end{equation}
The $4\times 4$ matrices  $K, \hat K$ are integrals over products
of Green's functions.  For notational convenience it is useful to
introduce an extra factor $i$ for the
$\sigma_2$-component in (\ref{Paulientw}).
 The structure of $K$ is given in the appendix
(\ref{K})
and the  functions are listed in (\ref{functions}).
With help
of these matrices $K$, $\hat K$ the vertex equations
can be written as
\begin{eqnarray}
(1 + W' K) \underline {\gamma} + \bar V \hat K
\underline {\hat \gamma}
& =& \underline {I}
\nonumber
\\
(1 + W' \hat K) \underline {\hat \gamma}  + \bar V K
\underline {\gamma} & =&
\underline {\hat I}
\label{second}
\end{eqnarray}

The matrix $\hat K$ differs  from the matrix $K$ only by the
sign  of  some  elements  in  the  first  row  and  column.  More
precisely: $\hat K = g K g$ with $g$ being the diagonal matrix
$g = diag(-1,  1, 1, 1)$.
$\underline{I}$ is a vector given by $(0,0,0,1)$, $(0,0,-i,0)$,
 $(0,1,0,0)$, $(-i,0,0,0)$ for $P^{13}$, $P^{12}$,$P^{11}$,
$P^{20}$, respectively.
$\underline{\hat I}$  fulfills:
$\underline{\hat I}=g\underline{I}$.
The
second equation  in (\ref{second}) can be reduced
to the first equation  by
setting
\begin{equation}
\underline {\hat \gamma} = g \underline {\gamma}
\label{gammahatgamma}
\end{equation}
and it is sufficient
to solve the first equation
\begin{equation}
\sum_j \left(\delta_{ij} + (W'   + \bar V g_{ii})
K_{ij}\right)\gamma_j = I_i
\label{vertexeqn}
\end{equation}
 which is a set of 4 linear equations.

A  closer  inspection   of  the  integrals
contained  in the matrix  $K$ shows that two of the integrals,
$K_{11}$ and $K_{22}$ are badly convergent.  They depend
logarithmically  on the  cut-off  in momentum space  in a similar
way as the integrals  in the self-consistency  equations
(\ref{divergent}).
These badly  convergent  integrals  can  be  eliminated  by  using
 relations (\ref{Hmp},\ref{Hmm}) which can be derived
from Ward identities.
Then the  cut-off is only needed in the calculation
of the gaps $\Delta_i$.

The polarization function of the neutral system  is  given by
\begin{eqnarray}
\lefteqn{
\ll P;P \gg_{i\omega_s} =}
\nonumber
\\
&& {1\over \beta} \sum_{\omega_n} \sum_k
Tr  \lbrace  D^{13}G(k,i\omega_n  +  i\omega_s)
\Gamma^{13}(i\omega_s) G(k,i\omega_n)\rbrace
\end{eqnarray}
The polarization function
 can be expressed by the matrix $B$ (\ref{B}) or by the
matrix $K$ using the commutation relations for the
Pauli matrices and
the development into Pauli matrices (\ref{Paulientw}):
\begin{equation}
\ll P,P\gg=\ll P^{13},P^{13}\gg=4 K_{3j}\gamma^{13}_j
\label{corrfkt}
\end{equation}

\subsection {Charged system}

Up to now we have only considered neutral superconductors.
The long-range Coulomb
interaction is known to have important consequences
for the collective modes
\cite{AndBogol,Schrieffer,Nambu,Parks}. Here we are interested in
the optical properties for field vectors perpendicular to the
layers  in the long-wavelength  limit $q \to 0$.  Then we have to
consider only the Coulomb interaction arising from charge
fluctuations between the layers.  These Coulomb forces stay finite
in the long-wavelength  limit but nevertheless are sizable.  They
can be incorporated in the vertex equations in a RPA-type manner.

\begin{figure}[h]
\begin{center}
\leavevmode
\epsfxsize=0.95\linewidth
\epsfbox{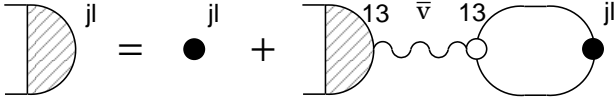}
\caption{\label{vertex3}
Diagrammatic representation of the vertex $\Upsilon^{jl}$  of the
charged  system  with  help  of the  vertex  $\Gamma^{jl}$  (full
circle) of the neutral system.  The interband Coulomb-interaction
$\bar   v$   couples   to   the   correlation    function    $\ll
P^{13},P^{jl}\gg $ of the neutral system. }
\end{center}
\end{figure}

Let $\Upsilon^{il}(i\omega_s)$ be the vertex for the
charged system. The vertex equation for $\Upsilon^{il}$
is   obtained   from   the   vertex   of   the   neutral   system
(\ref{vertexeqn1}),    if    $\Gamma^{jl}$    is   replaced    by
$\Upsilon^{il}(i\omega_s)$, and on the r.h.s.  of the vertex equation
(\ref{vertexeqn1}) the following term is included:
\begin{equation}
D^{13} \bar v \sum_{\omega_n}\sum_k\,\,
Tr\left\{
  D^{13} G(k,i\omega_n+i\omega_s)
\Upsilon^{jl}(i\omega_s)G(k,i\omega_n)
\right\}
\end{equation}
with $\bar v=\lim_{ |q|\to 0}e^2 \pi
( 1-{\rm e}^{-d|q|})/(|q|\epsilon_0)=e^2 \pi d/\epsilon_0$
  the Coulomb interaction between the layers.
For a periodic double-layer system $\bar v$ has
to be replaced by $\pi e^2 d(c-d)/c \epsilon_0$,
where $d$ is the distance between
the two layers in one unit cell and $c$ the distance
between the unit cells.
The vertex equation for $\Upsilon^{jl}$
can also be expressed by correlation functions and the vertex
of the neutral system  $\Gamma^{jl}$,
 as shown in Fig. \ref{vertex3}. This equation reads:
 \begin{equation}
\Upsilon^{jl}   =  \Gamma^{jl}   +  \Upsilon^{13}   \bar   v  \ll
P^{13},P^{jl}\gg
\end{equation}
where the correlation function  $\ll P^{13},P^{jl}\gg  $
of the neutral  system to which the interband
Coulomb-interaction couples is given by:
\begin{eqnarray}
\lefteqn{\ll P^{13},P^{jl}\gg = {1\over \beta} \sum_{\omega_n}
\sum_k}
\nonumber \\
&&Tr \lbrace D^{13}G(k,i\omega_n + i\omega_s) \Gamma^{jl}(i\omega_s)
G(k,i\omega_n)\rbrace
\end{eqnarray}

{}From this the vertex function $\Upsilon^{13}$ for the
electronic polarization is easily calculated:
\begin{equation}
\Upsilon^{13} = \Gamma^{13}/(1-\bar v \ll P^{13}, P^{13}\gg)
\label{Coulombin}
\end{equation}
and we get for the polarization function for the
charged system:
\begin{equation}
\ll P,P\gg_{i\omega_s}^C = \ll P,P \gg_{i\omega_s}
/\left(1-\bar v \ll P,P \gg_{i\omega_s} \right)
\label{PCPC}
\end{equation}

%%%%%%%%%%%%%%%%%%%%%%%%%%%%%%%%%%%%%%%%%%%%%%%%%%%%%%%%%%

\section{Analytical discussion of the collective modes}

In general the 4 vertex equations (\ref{vertexeqn})
\begin{equation}
\sum_j \left(\delta_{ij} + (W'   + \bar V g_{ii})
K_{ij}\right)\gamma_j = I_i \label{vertexeqna}
\end{equation}
have to be solved numerically.
But in some limits the
equations can be simplified so that analytical
results for the collective modes can be
derived:
 1.  in the normal state
   and   2.   in  the  limit   of  momentum   independent
tight-binding coupling $t_k$ between the layers.

In the following we need the special form of the matrix
$K$ given in the Appendix:
\begin{equation}
K = \left(
\begin{array}{cccc}
       H^{++} & -\tilde Y & -\tilde Z & \tilde X \\
              \tilde Y & - H^{-+} & X & Z \\
              \tilde Z & X & -H^{--} & Y \\
              \tilde X & -Z & -Y & H^{+-} \\
     \end{array}
\right)
\end{equation}
Furthermore we need the following
combinations of coupling constants, which are
listed here for completeness:
\begin{eqnarray}
W=V+J, \quad W_\parallel = V_\parallel + J, \quad
W_\perp = V_\perp - J \nonumber
\\
W'=V-J, \quad W'_\parallel = V_\parallel - J, \quad
W'_\perp = V_\perp - J
\end{eqnarray}
with $V=(V_\parallel + V_\perp)/2,
\quad \bar V=(V_\parallel + V_\perp)/2$.
 Of special importance is the limit, where all
pairing interactions are small
\begin{equation}
V N_0,\,\,\, \bar V N_0,\,\,\, J N_0 \ll 1
\label{weakcoupling2}
\end{equation}
This will be called the weak coupling limit in the following.

\subsection{Normal state}
 In the normal  state  the functions
$Z,\,\, \tilde Z,\,\, Y,\,\, \tilde Y$
vanish, because they are proportional to the gaps
(\ref{functions}).
Therefore the set of equations (\ref{vertexeqna}) decouples. With
$\underbar I = (0,0,0,1)$
we  obtain  for  the  polarization   vertex  $\gamma=\gamma^{13}=
\gamma^P$:
\begin{eqnarray}
\gamma_0 &=& - W^\prime_\perp \tilde X/N, \quad
\gamma_1=\gamma_2=0,
\nonumber \\
\gamma_3 &=&  (1  +  W^\prime_\perp
H^{++})/N  \end{eqnarray}
with $ N= (1 + W^\prime_\perp  H^{++})(1 +
W^\prime_\parallel  H^{+-})  - W^\prime_\parallel  W^\prime_\perp
\tilde X^2 $.
The polarization function (\ref{corrfkt}) is then given by
\begin{equation}
\ll P,P \gg_{i\omega_s}  = 4\left((1+ W^\prime_\perp H^{++})  H^{+-}
  - W^\prime_\perp (\tilde X)^2\right)/ N
\end{equation}

In the case of momentum independent  hopping ($t_k=t$)
this expression can be
simplified  considerably  by  using  the  relations  between  the
different  matrix elements derived from the Ward
identities (\ref{identities}) and the approximation
$(x_2-x_1)/W_\perp = - 2tN_0$. We obtain
\begin{equation}
\ll P;P \gg_{i\omega_s}  =
16N_0t^2(1+2N_0J)/M
\label{PPN}
\end{equation}
with
$$
M=(i\omega_s)^2 -4t^2[1-2(\bar V -2J)N_0 - 4J(\bar V -J)N_0^2]
$$
The polarization function has a simple pole which is shifted with
respect to the simple particle-hole excitation energy $2 t $
due  to the pairing  interactions.
In the weak  coupling  limit
(\ref{weakcoupling2})  we recover as result the bare
polarization  function,
which is given by $H^{+-}$.   In the charged  system there exists
then  a plasmon  describing  interlayer  charge  fluctuations  at
$\omega^2 = (2t)^2(1+\bar v N_0)$.

\subsection{$k$-independent hopping matrix-element $t$}

In the case of momentum  independent  tight-binding  coupling
($t_k=t$) the solution for the self-consistency
equations (\ref{xeqn},
\ref{gapeqn}) has the form
\begin{equation}
\Delta_1=\pm \Delta_2 \quad {\rm and\ } x_1=-x_2
\end{equation}
independent of the details of the band dispersion $\epsilon_k$.
Then there exists a symmetry relation between the band dispersions
$\xi_{1k}=\xi_k - t - x_1$ and $\xi_{2k}=\xi_k + t - x_2$
($\xi_k=\varepsilon_k - \mu$) and the quasi-particle
dispersions:
\begin{eqnarray}
&
   \xi_2(\xi_k)=-\xi_1(-\xi_k)&
\nonumber
\\
&
 E_1(-\xi_k)=E_2(\xi_k)
&
\label{xisymmetry}
\end{eqnarray}
Due to this symmetry some matrix functions $K_{ij}$ vanish
and therefore  (\ref{vertexeqna})  is reduced to a system of three
coupled  equations.   We  have  to  distinguish   the  two  cases
$\Delta_1= \pm \Delta_2$.

\subsubsection{Intralayer pairing  $\Delta_1=\Delta_2$}

In the case of dominant intralayer interaction $W_\parallel$ the
 band gaps are equal ($\Delta_1=\Delta_2=\Delta_\parallel$)
and interlayer pairing
vanishes ($\Delta_\perp = (\Delta_2-\Delta_1)/2 =0$).
Three  functions
are zero because of the relation (\ref{xisymmetry}):
\begin{equation}
Z(i\omega_s) = X (i\omega_s) = \tilde Y(i\omega_s) = 0
\end{equation}
 Therefore the second equation of the vertex equations
(\ref{vertexeqna}) can  be solved easily
\begin{eqnarray}
&
\gamma_1^j , \gamma_1^\Phi,\gamma_1^P = 0,\,\,
\gamma_1^A(i\omega_s)=1/\left(1-W_\parallel H^{-+}
(i\omega_s)\right),
&
\nonumber
\\
&
\gamma_l^A = 0,\,\,\, l =0,2,3
&
\label{gammaA}
\end{eqnarray}
This means that oscillations of the amplitude $A$
do not couple to oscillations of the current $j$,
polarization $P$, and phase
$\Phi$.  But the density, phase, and current oscillations
couple to each other. The Coulomb interaction does not affect the
amplitude fluctuations.
This is in agreement with the results found in  Refs.
\cite{WuGriffin,Kuboki},
where   a double layer system with $t=0$ is investigated.
In the case of k-dependent hopping all these quantities couple,
which will be
demonstrated below by our numerical results.

The frequencies of the collective modes are determined
by the zeros of the denominator
of the vertex functions $\gamma^{jl}$.
For the current, phase and polarization oscillations
the frequency of the collective
mode is given by  the determinant
of the corresponding $3\times 3$ matrix of the vertex-equation
system (\ref{vertexeqna}); the collective mode of
the amplitude oscillations is
determined by (\ref{gammaA}).

The functions
$K(\omega+i\delta)$ obtain imaginary parts at temperature $T=0$,
when the frequency  of the
external  field is large enough to break Cooper  pairs ($\omega  >
min(E_{1k}  + E_{2k})=\sqrt{(2\hat t\,)^2 +(2\Delta_\parallel)^2}$
with  $\hat t=t +(x_1-x_2)/2$).
We are interested in undamped collective modes below
the particle-hole threshold.
In order to derive a simple approximation
for the collective mode frequencies,
we shall assume small frequencies
$\omega\ll \sqrt{(2\hat t\,)^2 +(2\Delta_\parallel)^2}$,
a small hopping matrix element
$(\hat t/\Delta_\parallel)^2 \ll 1 $ and
zero temperature.

The functions $K$ (\ref{functions}) can be evaluated with help of
an expansion in $\hat t/\Delta_\parallel$ and
$\omega/2\Delta_\parallel$:
\begin{eqnarray}
&
 H^{+-}(\omega)\approx -N_0+\omega^2C,\,\,
 H^{++}    \approx (2\hat t\,)^2 C
&
\nonumber
\\
&
   Y(\omega) \approx \omega \left(N_0+((2\hat t\,)^2-\omega^2)
                         C\right)/(2\Delta_\parallel),
&
\nonumber
\\
&
   \tilde Z(\omega) \approx - 2\hat N_0/(2\Delta_\parallel),\,\,
   \tilde X \approx - \omega 2\hat t C
&
\label{approx}
\end{eqnarray}
where  $C=-N_0/(8 \Delta_\parallel^2)$.
Here $-(x_2-x_1)/(W_\perp 2 \hat t\,)$ has been
approximated by the density of states at
the Fermi level $N_0$ (see (\ref{x0})).

With these expressions we expand the determinant up to second order
in $\hat t/\Delta_\parallel$ or $\omega/(2\Delta_\parallel)$
\begin{equation}
Det=\frac{N_0 W_\parallel^\prime} {(2\Delta_\parallel)^2}
  (1-\omega_0^2 W_\parallel^\prime C)
 \left( \omega^2-\omega_{P}^2\right)
\end{equation}
Here $\omega_P$ is the frequency of
the collective phase mode in the neutral system
\begin{equation}
\omega_{P}^2 =
 \left[
       \omega_0^2 + (2\hat t\,)^2
    \left( 1 - W_\perp^\prime(N_0-\omega_0^2C) \right)
 \right]
     \frac{1-W_\parallel^\prime N_0}
{1-\omega_0^2 W_\parallel^\prime C}
\label{phaseparallel}
\end{equation}
$$
\mbox{with}\quad
\omega_0^2= \frac{(2\Delta_\parallel)^2} {N_0}\,
\frac{-2 J}{V_\parallel^2-J^2}
$$
This reduces to a simple result in
the weak coupling limit (\ref{weakcoupling2}):
\begin{equation}
\omega_{P}^2=(2t)^2 + \omega_0^2
\label{parallelwc}
\end{equation}
Because we have assumed small frequencies the formula
is only valid for small
$\omega_0$, i.e. $|J/V_\parallel|$ has to be
small.   The  expression  for the frequency  $\omega_0$  has been
derived  already by Leggett \cite{Leggett}.   Here we see how the
mode-frequency is modified by the interlayer hopping $t$.
It is justified to call this mode phase mode,
because it appears as a resonance in the
polarization function due to its coupling to the phase fluctuation.

Evaluating the polarization function (\ref{corrfkt})
of the neutral system
for small frequencies and small hopping matrix element
we obtain
\begin{equation}
\ll P,P\gg_\omega=
    4 N_0
\frac{\omega_{P}^2 ( 1-\omega_0^2 W_\parallel^\prime C)
      - \omega^2\,\omega_0^2 C\,(1-W_\parallel^\prime N_0)    }
     {\left(\omega^2-\omega_{P}^2\right)
       ( 1-\omega_0^2 W_\parallel^\prime C)
(1-W_\parallel^\prime N_0)}
\label{PPS}
\end{equation}

Inserting (\ref{PPS}) into  the polarization function  with Coulomb
interaction (\ref{PCPC})
we get the
phase-mode frequency in the charged system:
\begin{eqnarray}
\omega^2_{C}=
\left[\omega_0^2 +(2\hat t\,)^2
       \left(1-W_\perp^\prime(N_0-\omega_0^2 C)\right)
\right] \nonumber
\\
\times \frac{ 1+N_0(4\bar v - W_\parallel^\prime)}
     {1+\omega_0^2 C ( 4\bar v - W_\parallel^\prime)}
\label{phaseparallelC}
\end{eqnarray}
In the weak coupling limit (\ref{weakcoupling2}) and for small
$|\omega_0^2C|$ we have
\begin{equation}
\omega_C^2=(\omega_0^2+(2t)^2)(1+4N_0\bar v)
\label{phaseparallelCwc}
\end{equation}
The  interband Coulomb interaction $\bar v$ shifts
the mode up to higher frequencies.
$\bar v$ is proportional to the distance $d$ between the layers.
In order to get
a phase mode below the particle hole threshold,
it is necessary that
the distance of the layers
and the Josephson
coupling constant $J$ is small.
 If the Josephson  coupling $J$ is zero,
the resonance appears at $2t$ as in the
normal state but still below the particle-hole threshold.
The mode corresponds to the collective tunneling of Cooper pairs
between the layers without pair breaking.

The denominator of $\gamma_1^A$ (\ref{gammaA})
determines the amplitude mode.
With the relations (\ref{def1})
 and using Ward identities (\ref{identities}) we get:
\begin{equation}
\frac{1}{\gamma_1^A(\omega)} =
 1 - \frac{W_\parallel^\prime}{W_\parallel}-
  \left(\frac{(2\hat t\,)^2-\omega^2}{(2\Delta_\parallel)^2} +1\right)
   W_\parallel^\prime \frac{2\Delta_\parallel Y(\omega)}{\omega}
\end{equation}
For vanishing Josephson coupling
($W_\parallel^\prime/W_\parallel=1$) the
amplitude mode lies just at the particle-hole threshold:
$$
\omega_{A}^2=(2\hat t\,)^2+(2\Delta_\parallel)^2
$$
For finite Josephson coupling an approximate analytical result
can be obtained
with (\ref{approx}):
\begin{equation}
\omega_{A}^2=(2\hat t\,)^2+2(\omega_0^2 + (2\Delta_\parallel)^2)
\label{amplitudeparallel}
\end{equation}
Because we assumed small frequencies,
this result is only valid, if $\omega_0^2$ is negative,
which is obtained for
positive $J$. For the general solution we refer to
the numerical calculation.

The formula for the phase (\ref{parallelwc}) and amplitude mode
(\ref{amplitudeparallel})   show  how  the  mode frequencies  are
shifted    by   the   hopping    $t$   and   the    results    of
Refs.\cite{Leggett,WuGriffin} are modified. Note, however, that
the particle-hole  threshold  is shifted  too.

\subsubsection{Interlayer pairing $\Delta_1=-\Delta_2$}

In   the   case   of  dominant   interlayer   interaction,   when
$\Delta_\perp  = (\Delta_2-\Delta_1)/2\ne0,\,\,  \Delta_\parallel
=(\Delta_2+\Delta_1)/2  = 0$, the following  three functions  are
zero because of (\ref{xisymmetry}):
\begin{equation}
\tilde Z(i\omega_s) = X (i\omega_s) = Y(i\omega_s) = 0
\end{equation}
The third column and line of the vertex equation matrix
(\ref{vertexeqna}) is zero with the exception of the
diagonal element. Now the oscillations of $\Phi$
are decoupled from the
the oscillations of the other quantities, while $j$,
$ A$ and $P$ couple to each other.

Again we study the limit of small frequencies
$\omega\ll \sqrt{(2\hat t\,)^2
+\Delta_\perp^2}$ and small hopping $|\hat t/\Delta_\perp|^2\ll 1$.
The functions can be approximated as:
\begin{eqnarray}
&
  H^{++}\approx -N_0+\omega^2 C,
\,\,
  H^{+-}\approx (2\hat t\,)^2 C,
&
\nonumber
\\
&
  \tilde Y \approx\omega(N_0+((2\hat t\,)^2-\omega^2) C)
/(2\Delta_\perp),
&
\nonumber
\\
&
   Z\approx - 2\hat t N_0/(2\Delta_\perp),
\,\,
  \tilde X \approx - \omega 2\hat t C,
\nonumber
&
\\
&
  H^{-+}\approx 1/W_\perp -
(\omega^2-(2\hat t\,)^2) N_0/(2\Delta_\perp)
&
\end{eqnarray}
where $C=-N_0/8\Delta_\perp^2$ and $\hat t = t +(x_1-x_2)/2$
as before.
{}From  the  vanishing  of the  determinant  of the  three  coupled
equations we now obtain for the frequency of the collective mode
determining polarization and amplitude oscillations:
\begin{equation}
\omega_{P}^2 =
\left[ \omega_0^2
      +(2\hat t\,)^2 \left(1- W_\parallel^\prime (N_0-\omega_0^2 C )
\right)
\right]
\frac{1-W_\perp^\prime N_0}{ 1-\omega_0^2 W_\perp^\prime C}
\label{ampltitudeperp}
\end{equation}
with
$$
\omega_0^2= \frac{(2\Delta_\perp)^2}{N_0}\,
                          \frac{2(\bar V-J)}{V^2-(\bar V -J)^2}
$$
In the week coupling limit (\ref{weakcoupling2})
 the mode in the neutral
system becomes:
\begin{equation}
\omega_{P}^2=(2t)^2 + \omega_0^2
\label{amplitudeperpwc}
\end{equation}
Expanding the nominator and denominator of
the polarization function (\ref{corrfkt})
up to second order in $\hat t/\Delta_\perp$ and
$\omega/2\Delta_\perp$  we
obtain
\begin{equation}
\ll P,P\gg_\omega=
4 (N_0-\omega_0^2 C)
\frac{(2\hat t\,)^2(1- W_\perp^\prime N_0)}
     {\left(\omega^2-\omega_P^2\right)
       (1-\omega_0^2 W_\perp^\prime C) }
\label{PPS2}
\end{equation}
In this case the polarization function is zero,
if the single-particle hopping $t$ vanishes.
This is in contrast to the case of intralayer pairing,
where the pair tunneling due to $J$ produces  density  fluctuations
between  the  layers  and  therefore   spectral   weight  in  the
polarization  function  even  in the  absence  of single-particle
hopping.  P.~W.~Anderson   argues  that  due  to  correlations  the
single-particle hopping is suppressed.  If this is right, then in
the  case  of  interlayer   pairing   no  excitation   of  charge
fluctuations between the layers should be seen in the optical c-axes
conductivity   $\sigma(\omega)$,   which   is   related   to  the
polarization  function  by:   $\sigma(\omega+i\delta)  = i \omega
<<P,P>>_{\omega+i\delta} (ed/2)^2$.

The collective mode in the charged system is given by:
\begin{eqnarray}
\omega^2_C &=&
  \left[
   \omega_0^2 + (2\hat t\,)^2
           \left( 1 +(4\bar v - W_\parallel^\prime)
                    ( N_0 -\omega_0^2 C)\right)
  \right]
\nonumber
\\
&&\quad \times\,\,
\frac{1-W_\perp^\prime N_0}{1-\omega_0^2 W_\perp^\prime C}
\label{chargedmodeperp}
\end{eqnarray}
%
%}
Only the term containing the single-particle hopping is influenced
by the
Coulomb interaction. This is consistent with the picture of
interlayer pairing.
The hopping of interlayer Cooper-pairs
produces no charge fluctuation between
the layers. When an interlayer pair is tunneling,
 one electron jumps to the upper layer,
the other to the lower layer and
 the charge density on the layers does not change.

In a similar  way as above  we obtain  for  the frequency  of the
phase mode, which is now decoupled from the polarization:
\begin{equation}
\omega_{\Phi}^2 = (2\hat t\,)^2 + 2(\omega_0^2 + (2\Delta_\perp)^2)
\label{phaseperp}
\end{equation}
Because we assumed small frequencies,
this result is only valid, if $\omega_0^2$ is negative,
which is obtained for
negative $\bar V-J$.
 For the general solution we refer to the numerical calculation.

In comparison to the case of intralayer pairing
the modes have changed their role. The
amplitude  mode  is now the low lying  excitation  and is coupled
to the electronic polarization. As in the ground state
of interlayer pairing
 the layer order-parameters $<c_{1\uparrow} c_{1\downarrow}>$ and
$<c_{2\uparrow}  c_{2\downarrow}>$  are  zero  it  is  no  longer
justified to call oscillations of $\Phi$ phase oscillations.   We
kept this name for convenience only.

%%%%%%%%%%%%%%%%%%%%%%%%%%%%%%%%%%%%%%%%%%%%%%%%%%%%%%%%%%%%%%%%
\section{numerical results}
In the general case of k-dependent hopping
when intralayer and interlayer pairing are mixed it is not
possible to derive simple expressions
for the dispersions of the collective modes. We
made numerical calculations for the polarization functions
with the dispersion parameters (\ref{parameter}) and
coupling constants
\begin{equation}
N_0 W = -0.139, \quad N_0 \bar V = \pm 0.185, \quad N_0\bar v = 1.3
\label{Par}
\end{equation}
for constant $ W=V + J$ and variable Josephson coupling $J$
for dominating intralayer ($\bar V <0$) and interlayer
pairing ($\bar V >0$), in order to study the influence
of the Josephson
coupling $J$ and the temperature dependence of the collective modes.
We present results both for the
neutral  system and taking into account  the Coulomb  interaction
$\bar v$ between  the layers. To resolve the
$\delta$-peaks  of  the  collective  modes  at  $T=0$  below  the
particle-hole threshold we have introduced a small imaginary part
($10^{-3}$ meV) to the frequency.

\subsection{Dominant intralayer pairing}
\begin{figure}[h]
\begin{center}
\leavevmode
\epsfxsize=0.95\linewidth
\epsfbox{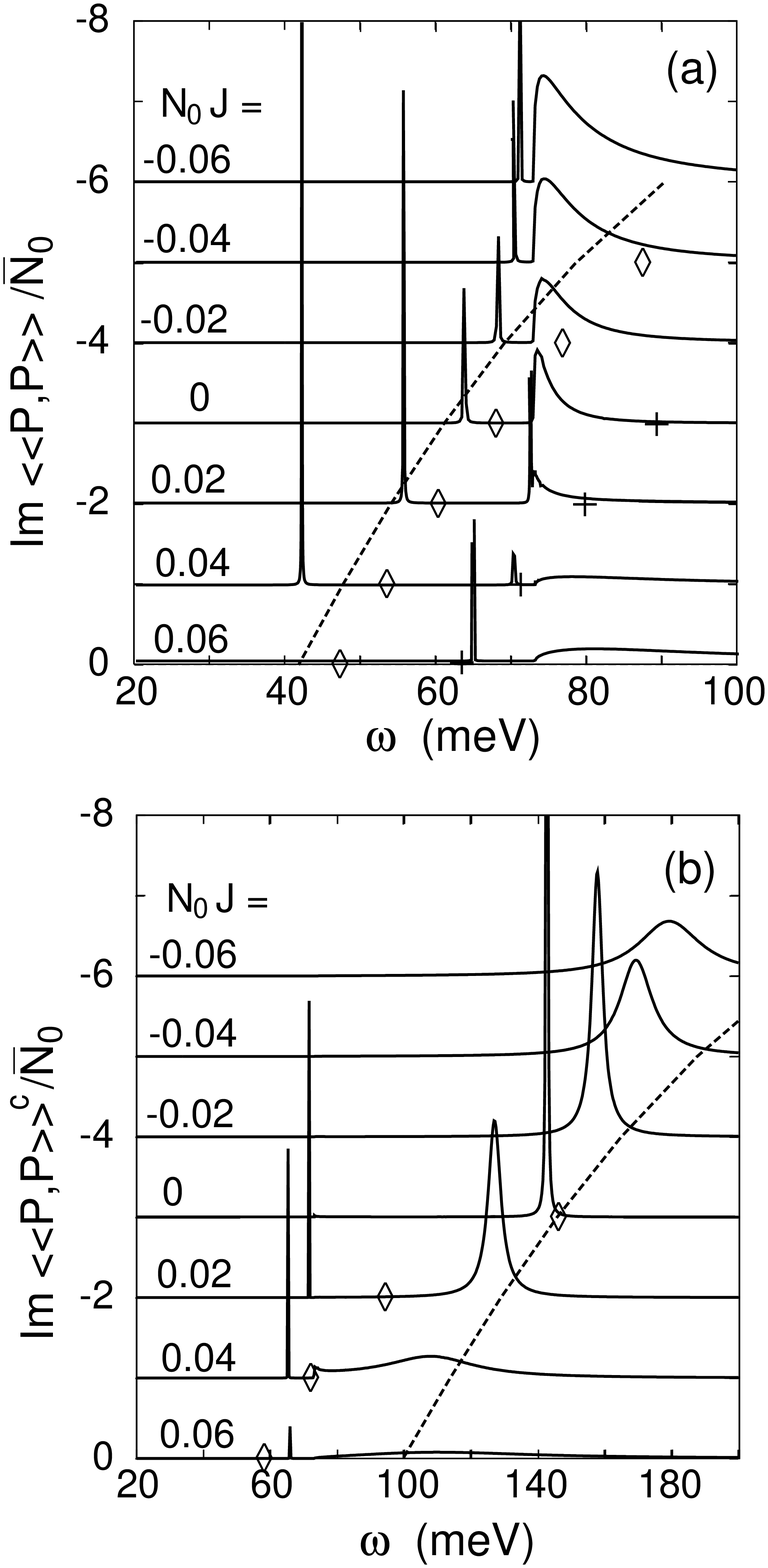}
\caption{\label{PPJI}Imaginary parts of the polarization function
in the neutral (a) and charged (b) system at $T=0$
in the case of dominant
intralayer  pairing for different Josephson couplings $J$.
The parameter sets (22,23) are used.  The results are normalized by
$\bar N_0 = 13 N_0$.  The different lines are shifted by equal units.
The rhombs,  crosses,  and dashed lines in the figures  correspond
to  various  approximation   formulas  for  the  collective  mode
frequencies discussed in the text.}
\end{center}
\end{figure}

First we discuss the results for the case of
dominant intralayer pairing ($\bar V  <0, \Delta_1 \Delta_2>0$).
In Fig. \ref{PPJI} the spectrum of the
collective modes (sharp spikes)
and the form of the particle-hole spectrum at $T=0$ is shown. The
imaginary  parts  of the  polarization  function  in the
neutral
\ref{PPJI}(a) and the charged system \ref{PPJI}(b) are calculated
for  different  Josephson  couplings  $J$ and constant  intraband
$W=V+J$ and interband  coupling $\bar V$.  For comparison
results  of the approximation  formulas  are plotted,  too.   The
rhombs  and crosses  in (a) refer  to the formulas  for the phase
(\ref{phaseparallel})          and         amplitude         mode
(\ref{amplitudeparallel}), respectively,   calculated   with  a
constant  averaged hopping $t=-30.5\,{\rm  meV}$.  The dashed line
represents  the phase mode frequency (\ref{parallelwc})
in the weak coupling  limit.

 For negative Josephson couplings $J$ a collective mode,
the phase mode, lies just
below the particle-hole threshold.
With increasing positive $J$ the phase-mode peak
moves to lower frequencies.

For positive $J$ another mode, the amplitude mode,
becomes visible near the
particle-hole threshold.
 For increasing  $J$ the peak moves to smaller frequencies. Thus, two
superconducting collective modes  can occur
in the polarization function
perpendicular to the layers. This effect is caused
by the coupling of the phase and
amplitude fluctuations.

The two approximation formulas for the phase mode,
represented by the rhombs and the
dashed line, give
qualitatively   the   right   position   of  the  collective-mode
frequencies,  as long  as the peaks  are  below  the  particle-hole
threshold.   The same holds for the approximated  values  for the
amplitude mode (crosses).

For $N_0 J\ge 0.06$ the peak of the phase mode has passed zero,
and its
frequency has become imaginary.
This behaviour is also obtained by using
the approximation formula (\ref{parallelwc}).
We believe, that this instability indicates a phase
transition from intraband to interband pairing.
This is supported by a study of a two-site model with
two electrons
(see section self-consistency equation), where a strong
positive coupling $J$ leads
to a ground state consisting of interband Cooper-pairs
($a^\dagger_{1 \uparrow} a^\dagger_{2\downarrow}$).

In Fig. \ref{PPJI}(b) the influence
of the  Coulomb interaction between the layers is shown.
A plasmon exists within the particle-hole continuum.
This mode is caused by charge
fluctuations between the layers. The damping of this mode
increases with increasing $|J|$.
The particle-hole threshold peak is strongly suppressed,
if there exists a collective mode
with large spectral weight above the threshold.
Again the approximation formula for the phase mode
in the weak coupling limit,
represented by the dashed line in Fig. \ref{PPJI}(b),
gives the right position of the plasmon peak
for small $|J|$. The plasmon mode and
phase mode coincide for small $|J|$.

But for positive $J$  a peak below the particle-hole threshold
appears ($N_0 J = 0.02,\,\,\, 0.04$).
This peak can be attributed to phase
fluctuations by comparing the peak positions
with the analytical results obtained for
constant $t$ (rhombs).
The amplitude peak lies just at the particle-hole threshold
at $N_0 J=0.04$ and
can be seen at $N_0 J=0.06$. Those
peaks which we have attributed to the phase and amplitude modes
have large spectral
weight in the correlations functions
$\ll \Phi,\Phi\gg$ and $\ll A,A\gg$, respectively.

Now we discuss the temperature dependence of the polarization
functions for  large positive (0.05)
 and negative (-0.06) Josephson
coupling.

\begin{figure}[h]
\begin{center}
\leavevmode
\epsfxsize=0.75\linewidth
\epsfbox{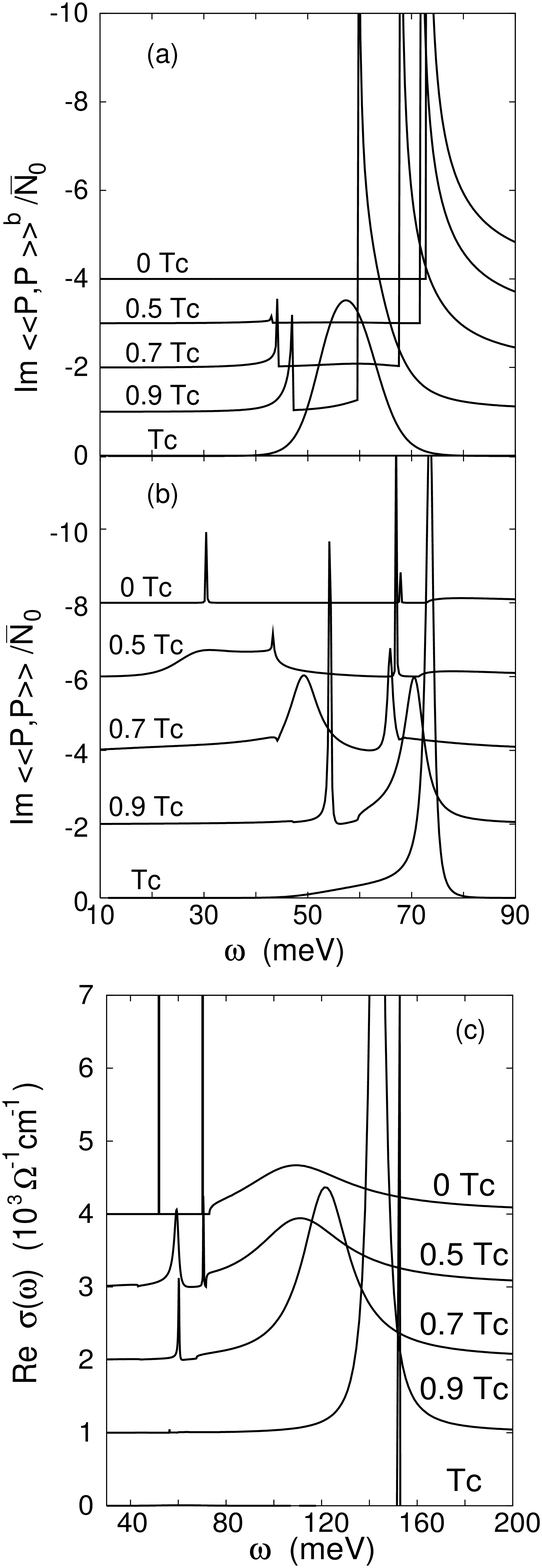}
\caption{\label{all}
Imaginary parts of the bare polarization function (a),
the polarization function in
the  neutral system (b), and  the real part of
the optical conductivity  in the charged system (c)
in the case of dominant  intralayer pairing
for different temperatures $T$ and positive
Josephson coupling $N_0 J=0.05$.
The  parameters (22,23) are used.
The different lines are shifted by equal units. In (c) we choose
as  distance between the layers $d=3  \AA$.}
\end{center}
\end{figure}

Fig. \ref{all} shows the imaginary part
of the bare polarization function (a), the
polarization function in the neutral system (b),
and the real part of the optical
conductivity in the charged system (c). The latter is related to
the polarization function by $\sigma(\omega+i\delta) = i \omega
(ed/2)^2\ll P,P\gg_{\omega+i\delta}^C$.
In picture \ref{all}(a) only particle-hole excitations
with energy $|E_{1k}+E_{2k}|$ are
visible at $T=0$. For finite temperatures also excitations
with $|E_{1k}-E_{2k}|$ from
excited levels are possible at lower frequencies.
The broad peak at $T_c$ is due to
interband transitions with different $t_k$-values.
The Fig. \ref{all}(b) shows the complicated transition
from two distinct collective
modes at $T=0$ (phase and amplitude mode)
to a collective density fluctuation in the
normal state. The Coulomb interaction
between the layers shifts the collective modes
associated with density fluctuations
to higher frequencies (Fig. \ref{all}(c)), whereas
the amplitude mode near the particle-hole threshold is only
slightly shifted.

\begin{figure}[h]
\begin{center}
\leavevmode
\epsfxsize=0.95\linewidth
\epsfbox{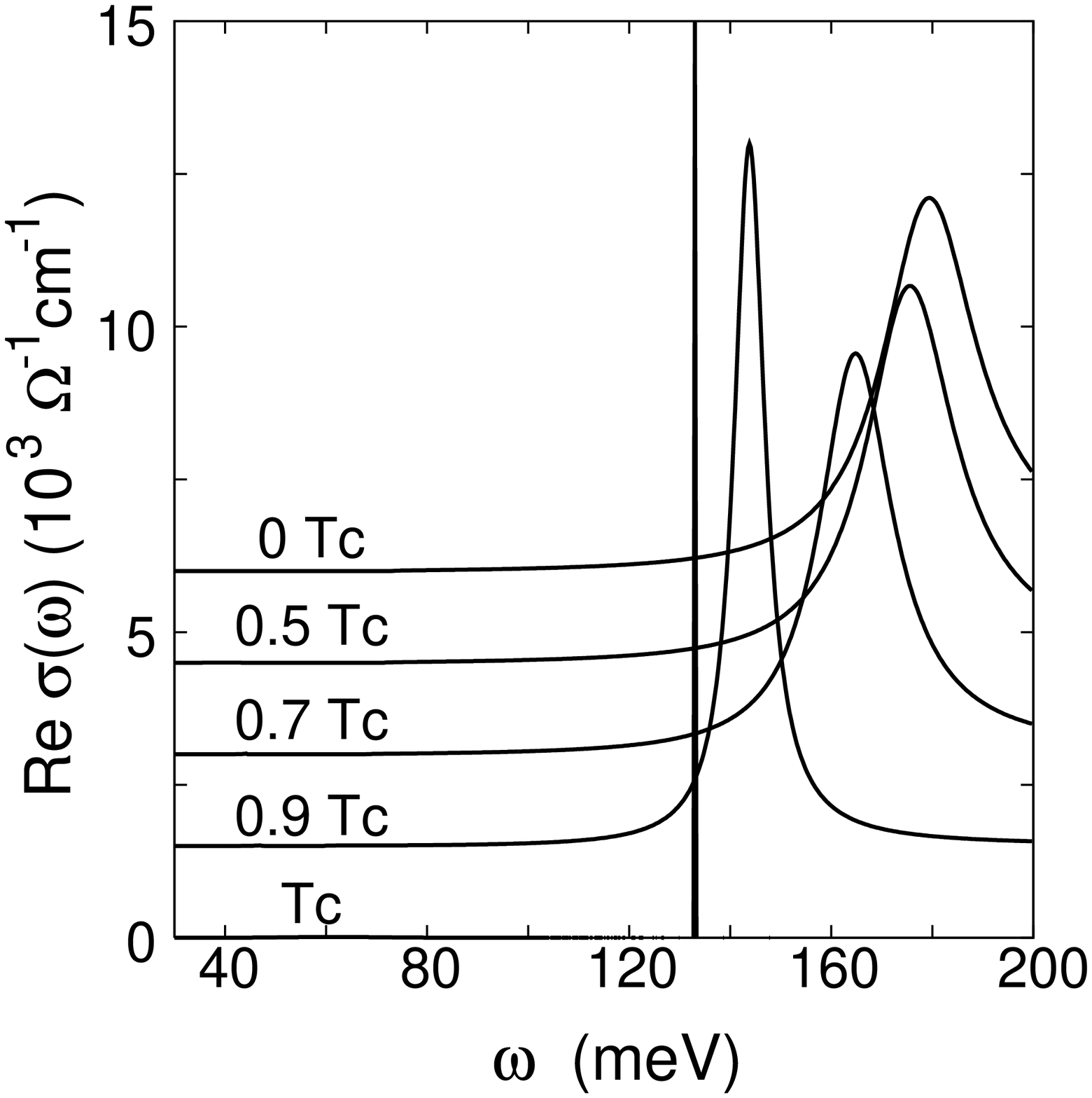}
\caption{\label{plasmon} Real part
of the optical conductivity $\sigma(\omega)$
in the case of dominant intralayer pairing
for different  temperatures $T$ and negative
Josephson coupling $N_0 J=-0.06$.
The parameters (22,23) are used, $d=3 \AA$.
The different lines are shifted by equal units.}
\end{center}
\end{figure}
In the case of negative $J$, which is the more realistic one,
if this coupling is produced by the exchange of a boson, the Coulomb
interaction shifts the collective modes
into the region of particle-hole excitations.
Fig. \ref{plasmon} shows the temperature dependence
of the remaining
density-fluctuation  spectrum.

\subsection{Dominant interlayer pairing}

\begin{figure}[h]
\begin{center}
\leavevmode
\epsfxsize=0.95\linewidth
\epsfbox{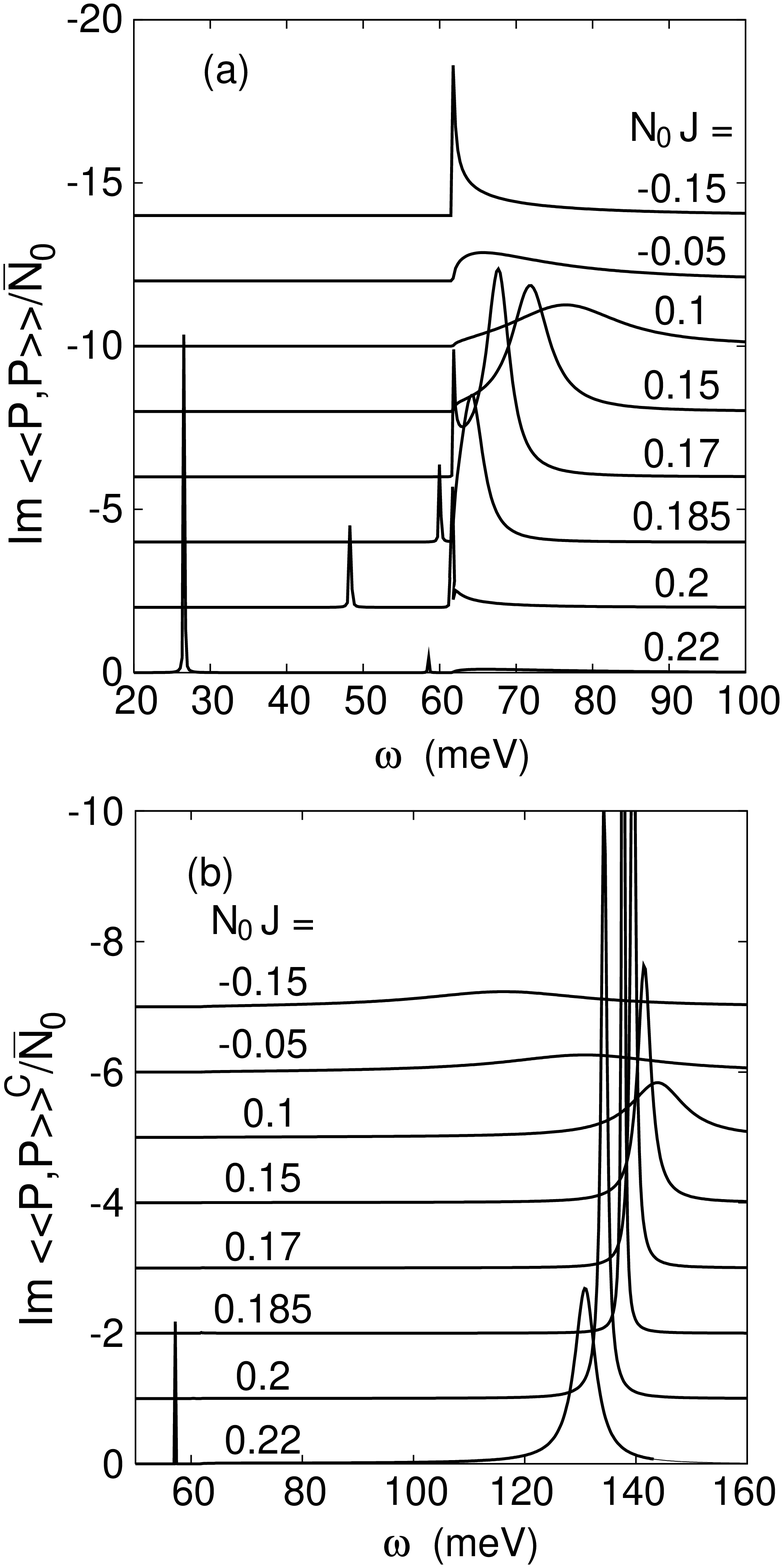}
\caption{\label{perp144}Imaginary part of  the
 polarization function in the neutral (a)  and charged (b) system
  in the case of dominant interlayer pairing
for different  Josephson couplings and temperature $T=0$.
The parameters (22,23) are used.
}
\end{center}
\end{figure}

In the case of dominant interlayer  pairing ($\bar V >0,
\Delta_1 \Delta_2 <0$) it is possible to produce
the same peak structures in
the polarization  function as in the case of dominant  intralayer
pairing.
Fig. \ref{perp144} shows the imaginary parts of
the polarization function in the neutral
(a) and charged  (b) system
for selected $J$-values.
The structures are nearly the same as in Fig. \ref{PPJI}.
For $J$ greater equal $\bar V$
collective modes below the particle-hole threshold can occur
in the polarization function of the neutral system
and for relative large positive $J$ also
in the polarization
function for the charged system. The damping behavior of the
plasmon is quite similar as in the case
of dominant intralayer pairing.
For negative and small positive $J$
the plasmon peak is strongly damped.

\begin{figure} [h]
\begin{center}
\leavevmode
\epsfxsize=0.95\linewidth
\epsfbox{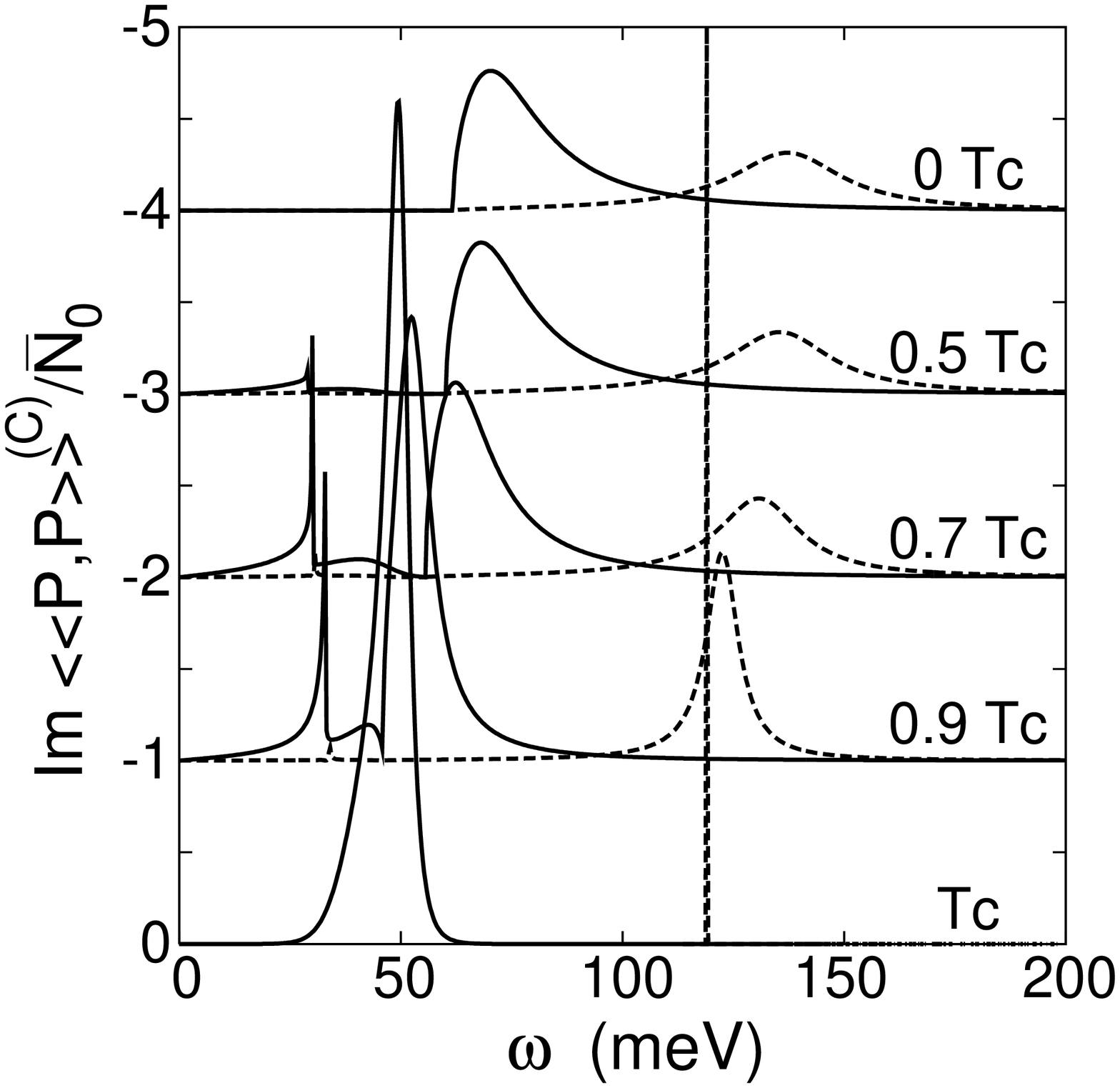}
\caption{\label{tempinterlayer}Imaginary part
of the polarization function in the
 neutral (full lines) and charged
(dashed lines)  system in the case of dominant interlayer
pairing
for different  temperatures and Josephson coupling $J=0$.
The  parameters (22,23) are used.
}
\end{center}
\end{figure}

An example for the temperature
dependence of the imaginary part of the
polarization function in the neutral (full lines)
and charged (dashed
lines) system shows Fig. \ref{tempinterlayer}. The plasmon (dashed
line) has the same behaviour as in Fig. \ref{plasmon}.

%%%%%%%%%%%%%%%%%%%%%%%%%%%%%%%%%%%%%%%%%%%%%%%%%%%%%%%%%%%%%%%%%%%%
\section{Conclusions}
In this paper we have studied collective modes in superconducting
double-layer systems  which  are  connected  with
charge  fluctuations
between  the layers.   In our model we assumed  a single-particle
tight-binding  hopping  $t_k$ between  the layers  and considered
three  different  types of pairing  interactions:   an intralayer
interaction  $V_\parallel$,  an interlayer interaction  $V_\perp$
and a Josephson-type  coupling $J$ which allows a transfer of two
particles between the layers. With these interactions we set up a
system of vertex equations within a conserving approximation.
The collective  modes  then  appear  as resonances  in the vertex
functions.  So we were able to study the interplay
of the three types of interlayer  interactions  $t_k$, $V_\perp$,
and $J$, which has not been investigated so far together.

Due to the interaction  $V_\perp$  Cooper pairs with electrons  in
different  layers are formed.  Depending  on the relative size of
the different  interactions  one finds dominating  interlayer  or
intralayer  pairing.   In the case  of constant  hopping  $t$  we
calculated analytically the frequency of the collective modes.
It is of
the  form  $\omega  = \sqrt{(2t)^2  + \omega_0^2}$ in the neutral
system.   In  the  case  of
intralayer  pairing  $\omega_0^2$  is proportional  to the  ratio
$-J/\vert V_\parallel\vert  $ while for interlayer  pairing it is
proportional  to
 $(V_\parallel-V_\perp -2 J)/\vert V_\parallel  + V_\perp\vert$
(in the limit that theses ratios are small).   The former result
is in agreement with the result for the collective
mode obtained by Wu and Griffin \cite{WuGriffin}  for $t=0$.  The
latter  result  is new.  In order to find an undamped  collective
mode  it is important  that its frequency  is below  the particle
hole  threshold  $\sqrt{(2t)^2+(2\Delta)^2}$.  As  for  realistic
systems  $\omega_0^2>0$  the best chances  to observe a collective
mode is for the case of intralayer  pairing  and small  Josephson
coupling  $J$.  However, one has to keep in mind that the Coulomb
interaction  between  the  layers  leads  to a further  shift  to
higher frequencies.  This Coulomb interaction is proportional  to
the  distance  between  the  layers  but can be reduced  by a large
phononic polarizability of the intermediate layers.

In principle one has to distinguish  two different modes
which  in the literature  sometimes are called  phase  and  amplitude
mode.  For $\omega_0^2<0$ the two collective modes,
can both have frequencies below the
particle-hole threshold for the
same coupling parameters  in the neutral system.  However, in the
case  of constant  $t$ and equal values of the two gaps
(pure intralayer or interlayer pairing)
only one of these  modes  couples  to the
electronic polarization and will show up in the optical spectra.
This is different for mixed intralayer and interlayer pairing
and finite hopping (this occurs in our model for k-dependent
hopping $t_k$).  Then both
phase and amplitude  oscillations  couple to the
charge oscillation  between the layers.  Therefore  in the c-axis
optical conductivity  two collective  mode peaks can appear below
the particle-hole threshold for low enough Coulomb interaction.

For a more realistic  parameter  region (negative  $J$ and strong
Coulomb  interaction)  only  one  collective  mode,  the  plasmon
corresponding  to charge fluctuations between the layers, exists.
In this case  most  of  the  weight  of
the  polarization   spectrum  is
concentrated  in the plasmon peak; the particle-hole  excitations
and  in  particular   the  peak at the threshold of particle-hole
excitations are  strongly suppressed.  Therefore it is difficult to
determine  the threshold  energy for breaking  up Cooper pairs or
the superconducting  gap with optical c-axis experiments.   It is
remarkable  that the plasmon peak is much narrower  in the normal
state  than  in the superconducting  state.   That  is due to the
stronger quasiparticle  dispersion in the superconducting  state,
allowing a much wider range of interband transition energies.

Up to now the  c-axis plasmon is not detected in the layered
high-$\rm T_c$ superconductors YBCO or BSCCO.
One explanation could be, that in these
compounds  the single-particle  hopping $t$ is zero (suppressed
by  correlation  effects,  as suggested  by P.W.   Anderson)  and
furthermore interlayer  pairing  dominates.
The  fluctuation  of  interlayer
Cooper-pairs  does not produce density  fluctuations  between the
layers.   The weight  of the polarization  function  would then only
be
determined by the hopping matrix-element  and therefore vanishes.
But it is also possible (not investigated  here) that the plasmon
is  overdamped  by  impurity  scattering  \cite{damping}  or more
likely by inelastic scattering effects at high frequencies due to
antiferromagnetic spin fluctuations.

\acknowledgments

This  work  has  been  supported  by  grants  from  the  Deutsche
Forschungsgemeinschaft (F.F.) and    the    Bayerische
Forschungsstiftung within the research project FORSUPRA (S.K.).

%%%%%%%%%%%%%%%%%%%%%%%%%%%%%%%%%%%%%%%
   \begin{appendix}
%%%%%%%%%%%%%%%%%%%%%%%%%%%%%%%%%%%%%%%%

%%%%%%%%%%%%%%%%%%%%%%%%%%%%%%%%%%%%%%%%%%%%%%%%%%%%%%%%%%%%

\section{Matrix Elements}
For the matrix $K_{ij}$ we have to calculate the following
integrals:
\begin{equation}
K_{ij}= {1\over\beta}\sum_{\omega_n k}
\frac{1}{2}Tr
 \left\{
  \bar\sigma_i^T\sigma_3
G_1(k,i\omega_n+i\omega_s) \bar \sigma_j  G_2(k,i\omega_n)
\sigma_3
\right\}
\label{Kij}
\end{equation}
where $\sigma_i$ are Pauli matrices and
the matrices  $\bar \sigma_j$ are defined by:
$$
\bar\sigma_j=\sigma_j\quad {\rm for\,\,} j=0,1,3,\quad
\bar\sigma_2=i \sigma_2,\quad\bar\sigma_2^T=-i \sigma_2
$$
Defining the integrals
$$
H^{\pm\pm}(i\omega_s) = {1\over \beta} \sum_{\omega_n}\sum_k
[(i\omega_n   +  i\omega_s)i\omega_n   \pm  \xi_{1k}\xi_{2k}   \pm
\Delta_1\Delta_2]/N
$$
$$
(X/\tilde X)(i\omega_s) = {1\over \beta} \sum_{\omega_n}\sum_k
[(i\omega_n   +  i\omega_s)\xi_{2k} \mp i\omega_n \xi_{1k}]/N
$$
$$
(Y/\tilde Y)(i\omega_s) = {1\over \beta} \sum_{\omega_n}\sum_k
[(i\omega_n   +  i\omega_s)\Delta_2 \mp i\omega_n \Delta_1]/N
$$
\begin{equation}
(Z/\tilde Z)(i\omega_s) = {1\over \beta} \sum_{\omega_n}\sum_k
[\xi_{1k} \Delta_2 \pm  \xi_{2k} \Delta_1]/N
\end{equation}
with
$$
N = [(i\omega_n + i\omega_s)^2 - E^2_{1k}][(i\omega_n )^2 -
                                           E^2_{2k}]
$$
The matrix $K$ can be written as
\begin{equation}
K = \left(
\begin{array}{cccc}
       H^{++} & -\tilde Y & -\tilde Z & \tilde X \\
              \tilde Y & - H^{-+} & X & Z \\
              \tilde Z & X & -H^{--} & Y \\
              \tilde X & -Z & -Y & H^{+-} \\
     \end{array}
\right)
\label{K}
\end{equation}
The different signs come from the commutation relations for the
Pauli matrices. Integrals of this type are common in vertex equations
for superconductors \cite{Marelexotic} and for
antiferromagnetic systems \cite{Brenig}.

The frequency summations are all of the general form
\begin{equation}
J(k, i\omega_s) = {1\over \beta} \sum_{\omega_n}
{F(i\omega_n + i\omega_s,i\omega_n)\over
((i\omega_n + i\omega_s)^2 - E_{1k}^2)((i\omega_n)^2 - E^2_{2k})}
\end{equation}
where the functions $F(z_1,z_2)$ are given by
\begin{eqnarray}
H^{\pm\pm}:    & F(z_1,z_2)&=    z_1z_2
\pm    \xi_{1k}\xi_{2k}    \pm \Delta_1\Delta_2
\nonumber
\\
(X/\tilde X): &F(z_1,z_2)&=    z_1 \xi_{2k} \mp z_2 \xi_{1k}
\nonumber
\\
(Y/\tilde Y): &F(z_1,z_2)&=    z_1 \Delta_2 \mp z_2 \Delta_1
\label{functions}
\\
(Z/\tilde Z): & F(z_1,z_2)&= \xi_{1k} \Delta_2 \pm \xi_{2k} \Delta_1
\nonumber
\end{eqnarray}
After performing the frequency summations  with help of Poisson's
summation formula we obtain:
\begin{eqnarray}
J(k,i\omega_s &)& =   {1\over   8E_{1k}E_{2k}}\left(\tanh   {\beta
E_{1k}\over 2} + \tanh   {\beta E_{2k}\over 2}\right)
\nonumber\\
&\times&\left({F(-E_{1k},E_{2k})\over i\omega_s + E_{1k}+E_{2k}} -
{F(E_{1k},-E_{2k})\over i\omega_s - E_{1k} - E_{2k}}\right)
\nonumber\\
&+& {1\over   8E_{1k}E_{2k}} \left(\tanh   {\beta
E_{1k}\over 2} - \tanh   {\beta E_{2k}\over 2}\right)
\\
&\times&\left({F(E_{1k},E_{2k})\over i\omega_s - E_{1k}+E_{2k}} -
{F(-E_{1k},-E_{2k})\over i\omega_s + E_{1k} - E_{2k}}\right)
\nonumber
\end{eqnarray}
One notices that the first term containing contributions from the
creation or destruction of two quasi-particles in different bands
remains  finite also for $T\to 0$.  The second term with transfer
of quasi-particles  from one band to the other  vanishes  in that
limit. Using the    expressions  for the function
$F(z',z)$  the different  matrix-elements  are easily calculated.
One observes  that the integrals  $H^{\pm\pm},  Z, \tilde  Z$ are
even functions of $\omega_s$  while the integrals  $X,\tilde X, Y,
\tilde Y$ are odd functions.

%%%%%%%%%%%%%%%%%%%%%%%%%%%%%%%%%%%%%%%%%%%%%%%%%%%%%%%%%%%%
\section{Vertex Functions and Ward Identities}

We derive some useful relations between different integrals
of the matrix $K$, which are based on the conserving
approximation used
for the calculation of self-energies and vertex-functions.
Similar relations for a one-band system are calculated e.g.
in Ref \cite{Varma}.

The vertex equation in the neutral system in ladder approximation
has the general form (\ref{vertexeqn1}).
\begin{equation}
\Gamma^{jl}(i\omega_s)     =    D^{jl} - L\lbrace \Gamma^{ij}\rbrace
\end{equation}
with
\begin{eqnarray}
\lefteqn{
L\lbrace \Gamma \rbrace ={1\over   \beta} \sum_{\omega_n k}
{\phantom{+}}
}\nonumber\\
&&{\phantom{+}}V   D^{03} G(k,i\omega_n+i\omega_s)\Gamma(i\omega_s)
                                 G(k,i\omega_n)  D^{03}
\nonumber \\
%&&{\phantom{{1\over   \beta} \sum_{\omega_n}\sum_k} }
&&+ \bar V D^{13}  G(k,i\omega_n+i\omega_s)
                \Gamma(i\omega_s)G(k,i\omega_n) D^{13}
\nonumber
\\
%&&{\phantom{{1\over   \beta} \sum_{\omega_n}\sum_k} }
&&+ J D^{33}  G(k,i\omega_n+i\omega_s)
                \Gamma(i\omega_s)G(k,i\omega_n) D^{33}
\end{eqnarray}
We consider only vertex functions $\Gamma$, which are off-diagonal
in the band indices:
\begin{equation}
\Gamma = \left(
      \begin{array}{cc}
                   0 & \gamma \\
              \hat \gamma &  0 \\
      \end{array}
     \right)
\end{equation}
Then $L$  is also  off-diagonal  in the  band indices  and  can  be
written as
\begin{equation}
L = \left(
      \begin{array}{cc}
                   0 & \lambda \\
              \hat \lambda &  0 \\
      \end{array}
     \right)
\end{equation}
Now let us define a k-dependent vertex function:
\begin{equation}
\Gamma^a_k(i\omega_s)
 = G^{-1}(k,i\omega_n + i\omega_s) D^{13} - D^{13}
G^{-1}(k,i\omega_n)
\end{equation}
then the right upper corner is:
\begin{equation}
\gamma^a_k=i\omega_s\sigma_3+2\hat t_k
\sigma_0 - i(\Delta_1+\Delta_2)\sigma_2
\end{equation}
where $\hat t_k = (\xi_{2k}-\xi_{1k})/2$.

Inserting  $\Gamma^a$ into  the  r.h.s. of (B2) some  of the
Green's function  cancel, and the integrals  are reduced to those
which also occur in the self-energies (\ref{sigma12}).
 In particular  we obtain
for the $2\times2$ matrix in the right upper corner
\begin{eqnarray*}
\lambda^a   &=& \sum_k\sum_{\omega_n}   \
  W'\left(  G_2(k,i\omega_n)\sigma_3  -  \sigma_3  G_1(k,i\omega_n+
                    i\omega_s)\right )
\\
&&{\phantom{\sum_k\sum_{\omega_n}}}
 + \bar V \left( G_1(k,i\omega_n)\sigma_3-\sigma_3
G_2(k,i\omega_n+i\omega_s)\right )
\\
&=&
\frac{1}{W^2-{\bar V}^2}
\\
&\times&
  \Big\{\,\,\,
         W'\left[
                 \sigma_3(-W \Sigma_2 + \bar V \Sigma_1)
                 +(W\Sigma_1 -\bar V \Sigma_2)\sigma_3
           \right]
\\
&&
       +\bar V\left[
                    \sigma_3(-W\Sigma_1+\bar V\Sigma_2)
                  + (W\Sigma_2-\bar V \Sigma_1)\sigma_3
              \right]
  \Big\}
\\
\end{eqnarray*}
\begin{equation}
=\frac{W'+\bar V}{W+\bar V}( \Delta_2+\Delta_1) i\sigma_2+
\frac{W'-\bar V}{W-\bar V}(x_2-x_1)\sigma_0
\end{equation}

On the other hand we can calculate $L\{\Gamma\}$ in an other way.
With the quantities $B$ and $\hat B$ defined in (\ref{B}) $\lambda$
is given by:
\begin{equation}
\lambda^a = W^\prime B + \bar V \hat B
\end{equation}
Because of the $k$-dependence of the vertex function $\gamma^a_k$
we introduce the $k$-dependent $4\times4$ matrix functions $K_k$
and $\hat K_k$, which are related to $K$ and $\hat K$ by
$$
K=\sum_k K_k,\quad \hat K = \sum_k \hat K_k
$$
The same relations we needed for deriving the vertex equation
(\ref{vertexeqna}) holds for the $k$-dependent quantities $K_k$,
$\hat K_k$, $\gamma^a_k$, and $\hat \gamma^a_k$. Therefore we find:
\begin{equation}
\underline \lambda^a = (W^\prime+g\bar V)\sum_k K_k
               \underline\gamma^a_k
\end{equation}
where $g$ is the diagonal matrix $g=diag(-1,1,1,1)$.
Thus we arrive at the first identity:
\begin{equation}
(W+g\bar V)\sum_ k K_k\underline\gamma^a_k  =
(x_2-x_1,0,\Delta_2+\Delta_1,0)^t
\end{equation}
Note on the l.h.s. stands $W$ not $W'$, because the factor
$W'+g\bar V$
cancels.

In a similar way we find a second relation starting from
\begin{equation}
\Gamma_k^b(i\omega_s)
 = G^{-1}(k,i\omega_n + i\omega_s) D^{10} - D^{10}
                                    G^{-1}(k,i\omega_n)
\end{equation}
Inserting this into the integral $L$ we obtain
\begin{eqnarray*}
\lambda^b  & =&  \sum_k\sum_{\omega_n}
  W' \sigma_3 G_2(k,i\omega_n) \sigma_3
 - \bar V \sigma_3 G_1(k,i\omega_n)  \sigma_3
\\
&&
-  W'  \sigma_3  G_1(k,i\omega_n+  i\omega_s) \sigma_3 +
\bar  V \sigma_3  G_2(k,i\omega_n+ i\omega_s)\sigma_3
\end{eqnarray*}
\begin{equation}
= {W' + \bar V \over W - \bar V} (\Sigma_1 - \Sigma_2)
\end{equation}
or
\begin{equation}
\sum _k K_k \underline \gamma_k^b= {1\over W - \bar V}
(0, (\Delta_2-\Delta_1),0, - (x_2-x_1) )^t
\end{equation}

In the case of vanishing Josephson coupling and k-independent
hopping $t$ the first relation is identical with the Ward identity
\begin{equation}
i\omega_s\Gamma^P + 2it\Gamma^j = \Gamma^a_k
\end{equation}
which follows from the continuity equation $2t j = \dot P$
(compare the discussion to (\ref{operators})). $\Gamma^P$ is here
the polarization vertex and $\Gamma^j$ the current vertex.
The second relation  is not  a Ward identity based on
a conservation law.
It is valid only within the ladder approximation  for
the vertex equation.
More relations  are obtained by replacing  the matrix $D^{13}$ by
$D^{11}$  or $D^{12}$.   These  relations,  however,  are not  of
practical use.

{}From the above relations we obtain the following identities
between  the different  integrals:
\begin{equation}
\begin{array}{ccccccc}
2 H^{t++}& +&\tilde Z 2 \Delta_\parallel&  +&
\tilde X i\omega_s&  =& (x_2-x_1)/W_\perp  \\
2 \tilde Y^t & -& X 2\Delta_\parallel & +&
Z i\omega_s&   =& 0 \\
2 \tilde Z^t &+& H^{--}2\Delta_\parallel&  +&
Yi\omega_s&   =& 2\Delta_\parallel /W_\parallel  \\
2 \tilde X^t& +& Y 2\Delta_\parallel &  +&
H^{+-}i\omega_s&  =& 0 \\
H^{++} i\omega_s  &+& \tilde Y 2\Delta_\perp & + &
2 \tilde X^t&  =& 0\\
\tilde Y  i\omega_s & +& H^{-+}  2\Delta_\perp &  + &2 Z^t&
= &2\Delta_\perp/W_\perp  \\
\tilde Z i\omega_s & -& X 2\Delta_\perp  &+& 2 Y^t & =& 0 \\
\tilde X i\omega_s& +& Z 2\Delta_\perp & + &
2 H^{t+-} &=& (x_2-x_1)/W_\perp
\end{array}
\label{identities}
\end{equation}
where $W_\parallel=V_\parallel+J = W+\bar V$,
$W_\perp=V_\perp+J=W-\bar V$ and $\Delta_\parallel = (\Delta_2 +
\Delta_1)/2, \Delta_\perp = (\Delta_2 - \Delta_1)/2$.
Here the quantities  $K^t_{ij}$  are integrals  similar to $K_{ij}$
with the integrand multiplied by $\hat t_k = t_k + (x_1 - x_2)/2$
before  the  momentum  integration.   In  the  case  of  momentum
independent  hopping  matrix-element  $t_k$  we have
$K^t_{ij} = \hat t K_{ij}$.

The identities  can be used    to
eliminate the badly converging integrals $H^{-+}$ and $H^{--}$:
\begin{eqnarray}
\label{Hmp}
  H^{-+}  &=& \frac{1}{W_\perp}   -
{1\over \Delta_2 - \Delta_1 } (i \omega_s \tilde Y+ 2 Z^t)
\\
\label{Hmm}
H^{--} &=& \frac{1}{W_\parallel} -
{1\over  \Delta_1 + \Delta_2 } (i \omega_s Y  + 2 \tilde Z^t)
\end{eqnarray}
 Here  the integrals  on the
r.h.s.  do not need a cut-off.

In the two special cases of pure intralayer and interlayer pairing
one  can derive from the definitions of the functions
(\ref{functions}) two further relations:
\begin{equation}
H^{-+} = H^{--} + (\Delta_1+\Delta_2) Y/i\omega_s
                     \quad\mbox{if\ }\Delta_1=\Delta_2
\label{def1}
\end{equation}
\begin{equation}
H^{--} = H^{-+} + (\Delta_2-\Delta_1) \tilde Y/i\omega_s
                       \quad\mbox{if\ }\Delta_1=-\Delta_2
\label{def2}
\end{equation}

%%%%%%%%%%%%%%%%%%%%%%%%%%%%%%%%%%%%%%
   \end{appendix}
%%%%%%%%%%%%%%%%%%%%%%%%%%%%%%%%%%%%%%

%%%%%%%%%%%%%%%%%%%%%%%%%%%%%%%%%%%%%%%%%%%%%%%%%%%%%%%%%%%%

\end{document}